\documentclass[a4paper,12pt]{article}
\usepackage{amssymb,amsmath,bm}
\usepackage{graphicx}
\usepackage{enumerate}
\usepackage{mathrsfs}
\usepackage{amsthm}

\title{Non-Smooth Integrability Theory\thanks{At the 2022 Fall Meeting of the Japan Economic Association, the author received helpful comments and suggestions from Hisatoshi Tanaka. The author is grateful to him. The author is also grateful to an associate editor for his/her kind comments and suggestions. This work was supported by JSPS KAKENHI Grant Number JP21K01403.}}
\author{Yuhki Hosoya\thanks{E-mail: ukki(at)gs.econ.keio.ac.jp,\ ORCID ID:0000-0002-8581-4518}~{}\thanks{Faculty of Economics, Chuo University. 742-1, Higashinakano, Hachioji, Tokyo, 192-0393 Japan.}}

\begin{document}
\maketitle

\begin{abstract}
We study a method for calculating the utility function from a candidate of a demand function that is not differentiable, but is locally Lipschitz. Using this method, we obtain two new necessary and sufficient conditions for a candidate of a demand function to be a demand function. The first concerns the Slutsky matrix, and the second is the existence of a concave solution to a partial differential equation. Moreover, we show that the upper semi-continuous weak order that corresponds to the demand function is unique, and that this weak order is represented by our calculated utility function. We provide applications of these results to econometric theory. First, we show that, under several requirements, if a sequence of demand functions converges to some function with respect to the metric of compact convergence, then the limit is also a demand function. Second, the space of demand functions that have uniform Lipschitz constants on any compact set is compact under the above metric. Third, the mapping from a demand function to the calculated utility function becomes continuous. We also show a similar result on the topology of pointwise convergence.

\vspace{12pt}

\vspace{12pt}
\end{abstract}

\section{Introduction}
In classical economics, measuring utility was important because the sum of utilities was seen as the most important index for determining the goodness of the society. After the latter half of the 19th century, it gradually became clear that this measuring problem was very difficult, and economics placed the greatest emphasis on analyses in which utility need not be measured (the so-called axiomatic approach). However, once Debreu (1974) had proved the Sonnenschein--Mantel--Debreu theorem, the limitations of the axiomatic approach became clear. That is, we can determine almost no properties of the economy without a specification of the utility functions.

Therefore, a method of estimating utility functions from data has become necessary. However, an important problem arises here. Namely, the only data available for estimating utility functions are those relating to the consumer's purchase behavior, and thus, only the demand function, not the utility function, can be directly estimated from data. Because of this, the estimation method currently used is basically a method of parameter estimation in which the demand function and the utility function are exogenously assumed to correspond one-to-one through their parameters. This method is known as calibration.

Our purpose is to construct a general theory that calculates the utility function from an arbitrary demand function without such an exogenous assumption. This research area is known as integrability theory. In this context, Hosoya (2017) developed a specific method for calculating the corresponding utility function from a given candidate of the demand function. This previous paper also discussed how to handle cases in which the demand function contains errors. Let us elaborate on this issue. To obtain the demand function, we need to estimate it from the purchase behavior of the consumer, as discussed above. The problem addressed by Hosoya (2017) is the following: if the error in the demand function is small, is the corresponding error in the utility function also small?

The problem is essentially that of continuity. In other words, it requires the property that, when the demand function changes slightly, the corresponding utility function also changes only slightly. The most important aspect of the continuity problem is the topology. Because Hosoya (2017) assumed that the demand function is continuously differentiable, the local $C^1$ topology was used for the space of demand functions. However, this result is problematic in many ways. The most significant problem is that there are few econometric results that discuss the space of the demand function with the local $C^1$ topology. This is because there are very few estimation methods that allow the error to converge to $0$ with respect to the local $C^1$ topology as the data size increases. Therefore, even if good results are obtained with respect to this topology, the results cannot be used in econometrics.

For the above reason, a weaker topology is needed. Common topologies used in econometric theory for such problems are the topology of compact convergence and the topology of pointwise convergence. However, using these topologies means that the space of continuously differentiable functions becomes not complete. We therefore need to reproduce the results of Hosoya (2017) without assuming that the demand function is differentiable. This is the research objective of the present paper.

Weakening the assumption of differentiability to that of continuity, however, means that we are confronted with the problem found by Mas-Colell (1977). He demonstrated the existence of a continuous demand function such that there are two corresponding continuous utility functions representing different preference relations. Applying this to our context means that, in the space of continuous demand functions, even if the error of the demand function is $0$, the estimated error of the utility function may not be $0$. This is highly undesirable.

Thus, as a compromise, we assume that the demand function is locally Lipschitz. By Rademacher's theorem, any locally Lipschitz function is differentiable almost everywhere, and thus, for a locally Lipschitz candidate of the demand function, we can define the Slutsky matrix almost everywhere. In this paper, we first use this result to extend Theorem 1 of Hosoya (2017). Namely, when a candidate of the demand function satisfies Walras' law and is locally Lipschitz, if its Slutsky matrix is symmetric and negative semi-definite almost everywhere, then the corresponding utility function can be constructed by solving a differential equation (Proposition 1). We present an example of such a calculation where this method works effectively (Example 1).

This result has several important consequences. First, using this result, we can obtain two conditions that are each necessary and sufficient for a candidate of the demand function to be a demand function (Corollary 1). As was the case under continuous differentiability, the symmetry and negative semi-definiteness of the Slutsky matrix ``almost everywhere'' constitute one necessary and sufficient condition. A more important necessary and sufficient condition is the existence of a unique global concave solution to a specific partial differential equation with an arbitrary initial value condition. This property is robust in terms of limit manipulation, i.e., it holds for the limit of a sequence of functions satisfying it. Hence, it provides an important stepping stone for the subsequent arguments in this paper.

The utility function corresponding to a demand function obtained by our method is upper semi-continuous on the range of the demand function, and we guarantee the uniqueness of the corresponding upper semi-continuous preference relation on the same space. Outside of the range of the demand function, Hosoya (2020) derived a construction method for an upper semi-continuous utility function and ensured the uniqueness of the corresponding upper semi-continuous preference relation when the range of the demand function is sufficiently wide. We present a slight variation of this construction method and prove that, again, if the range of the demand function is sufficiently wide, an upper semi-continuous utility function can be constructed outside the range of the demand function and the corresponding upper semi-continuous preference relation is unique (Corollary 2). Thus, it is not possible to obtain under our assumptions the non-uniqueness examples that Mas-Colell (1977) obtained for continuous demand functions.

With these results as the groundwork, we finally discuss the main focus of this study, namely the continuity of the mapping from the demand function to the utility function. As already mentioned, one of the typical topologies that can be given to the space of demand functions is the topology corresponding to uniform convergence on compact sets. In this space, even if a sequence of locally Lipschitz demand functions converges to some function (which is not necessarily a demand function), it is not guaranteed that the limit is locally Lipschitz. However, if the limit happens to be locally Lipschitz, we can show that the limit of a sequence of demand functions is still a demand function (Theorem 1).

With this in mind, we construct a certain space of functions. Specifically, this is the space of demand functions that satisfy Walras' law and have a uniform Lipschitz constant on any compact set. Using Theorem 1, we can prove the compactness of this space (Corollary 3). That is, in this space, every sequence of demand functions has a convergent subsequence whose limit is also a demand function.

For a unique derivation of the utility function, however, the range of the demand function must be sufficiently wide, as discussed above. We have found an example of a sequence of demand functions that satisfy all the assumptions of Corollary 3 and have a sufficiently wide range, yet the range in the limit is very small (Example 2). Therefore, an additional assumption is needed for the continuity result we desire. Namely, the range of the function in the limit must also be sufficiently wide. In addition, when all functions satisfy the ``C axiom'' introduced by Hosoya (2017, 2020), then the desired continuity proposition can be obtained (Theorem 2). That is, when a sequence of demand functions converges to a demand function with respect to the topology discussed above, then the corresponding sequence of utility functions also converges to the corresponding utility function uniformly on any compact set consisting of strictly positive consumption vectors.

Note that this result does not hold at the boundary: that is, when some commodity can be zero, we can only derive a more naive result (Corollary 4). Actually, we construct an example in which the sequence of values of the utility functions does not converge to the value of the limit utility function at the boundary (Example 3).

By strengthening the C axiom, we can strengthen the result of Corollary 3 and Theorem 2. We construct a new space of demand functions such that the C axiom is uniformly imposed on the whole space. We can then show that this space is compact, and furthermore, the mapping from this space to the space of the corresponding utility functions is continuous (Corollary 5). This is our desired result.

These are all results for the topology of compact convergence. We derive the same result as Corollary 5 for the topology of pointwise convergence. That is, if a sequence of demand functions in the space treated in Corollary 5 converges to some function pointwise, then this limit is also a demand function belonging in the same space, and the corresponding sequence of utility functions converges uniformly to the utility function corresponding to the limit demand function (Theorem 3). This is another desired result.

The results in the second half of the paper are specifically constructed with a view to discussing the consistency of estimation methods. An estimation method of the true value $x$ is said to be consistent if the estimated value $x_N$ converges in probability to $x$ as the data size $N$ increases. To summarize our results, we can state the following: if the estimation method for the demand function satisfies consistency, then the estimation method for the utility function that is constructed by the given estimation method for the demand function and the computational process in Proposition 1 also satisfies consistency. We believe that this presents a new way of estimating utility functions for econometric theory. In particular, this result can be applied to any estimation method, whether parametric or non-parametric.

The structure of this paper is as follows. First, Section 2 defines several terms in consumer theory that are necessary for understanding this paper. Section 3 introduces a method of constructing the utility function. Section 4 discusses the compactness of the space and the continuity of the representation results. Section 5 considers the position of the present work in the context of related research and provides a list of open problems. Section 6 summarizes the conclusions of this study. Because many of the theorems in this paper have long proofs, all proofs are placed in Section 7.

\section{Notation and Definitions}
Throughout this paper, we use the following notation: $\mathbb{R}^N_+=\{x\in\mathbb{R}^N|x_i\ge 0\mbox{ for all }i\in \{1,...,N\}\}$, and $\mathbb{R}^N_{++}=\{x\in \mathbb{R}^N|x_i>0\mbox{ for all }i\in \{1,...,N\}\}$. The former set is called the \textbf{nonnegative orthant} and the latter set is called the \textbf{positive orthant}. We write $x\ge y$ if $x-y\in \mathbb{R}^N_+$ and $x\gg y$ if $x-y\in \mathbb{R}^N_{++}$. If $N=1$, then we omit $N$ and simply write $\mathbb{R}_+$ and $\mathbb{R}_{++}$.

Fix $n\ge 2$. Let $\Omega$ denote the consumption set. We assume that $\Omega=\mathbb{R}^n_+$ unless otherwise stated. A set $A\subset \Omega^2$ is called a {\bf binary relation} on $\Omega$.

For a binary relation $A\subset \Omega^2$, we say that it is
\begin{itemize}
\item {\bf complete} if, for every $(x,y)\in \Omega^2$, either $(x,y)\in A$ or $(y,x)\in A$,

\item {\bf transitive} if $(x,y)\in A$ and $(y,z)\in A$ imply $(x,z)\in A$,

\item {\bf upper semi-continuous} if, for every $x\in \Omega$, the set $U(x)=\{y\in \Omega|(y,x)\in A\}$ is closed,

\item {\bf continuous} if $A$ is closed in $\Omega^2$,

\item {\bf upper semi-continuous on $B$} if, for every $x\in B$, the set $U^B(x)=\{y\in B|(y,x)\in A\}$ is closed with respect to the relative topology of $B$, and

\item {\bf continuous on $B$} if $A\cap B^2$ is closed in $B^2$.
\end{itemize}

A binary relation $\succsim$ on $\Omega$ is called a {\bf weak order} if it is complete and transitive. For a weak order $\succsim$, we write $x\succsim y$ instead of $(x,y)\in \succsim$ and $x\not\succsim y$ instead of $(x,y)\notin \succsim$. Moreover, we write $x\succ y$ if $x\succsim y$ and $y\not\succsim x$, and $x\sim y$ if $x\succsim y$ and $y\succsim x$.

Suppose that $\succsim$ is a weak order on $\Omega$. If there exists a function $u:\Omega\to \mathbb{R}$ such that
\[x\succsim y\Leftrightarrow u(x)\ge u(y),\]
then we say that $u$ {\bf represents} $\succsim$, or $u$ is a {\bf utility function} of $\succsim$.

Consider a function $f:\mathbb{R}^n_{++}\times\mathbb{R}_{++}\to \Omega$. We call the following condition the {\bf budget inequality}:
\[p\cdot f(p,m)\le m.\]
If the budget inequality holds for all $(p,m)\in \mathbb{R}^n_{++}\times \mathbb{R}_{++}$, then we call $f$ a {\bf candidate of demand} (CoD). Moreover, if
\[p\cdot f(p,m)=m\]
for all $(p,m)\in \mathbb{R}^n_{++}\times \mathbb{R}_{++}$, then we say that this CoD $f$ satisfies {\bf Walras' law}.

Let $\succsim$ be a weak order on $\Omega$. For each $(p,m)\in \mathbb{R}^n_{++}\times \mathbb{R}_{++}$, we define
\[\Delta(p,m)=\{x\in \Omega|p\cdot x\le m\},\]
\[f^{\succsim}(p,m)=\{x\in \Delta(p,m)|x\succsim y\mbox{ for all }y\in \Delta(p,m)\}.\]
We call the set-valued function $f^{\succsim}$ the {\bf demand relation} of $\succsim$, and if it is single-valued, then we call $f^{\succsim}$ the {\bf demand function} of $\succsim$. If $u$ represents $\succsim$, then $f^u$ denotes $f^{\succsim}$. For a CoD $f$, if $f=f^{\succsim}$, then we say that $f$ corresponds to $\succsim$ and $\succsim$ corresponds to $f$. Of course, if $f=f^u$, then we say that $f$ corresponds to $u$ and $u$ corresponds to $f$. We call a CoD $f$ a {\bf demand function} if $f=f^{\succsim}$ for some weak order $\succsim$ on $\Omega$.

Let $f$ be a CoD. We say that $f$ is {\bf income-Lipschitzian} if for every compact subset $C\subset \mathbb{R}^n_{++}\times \mathbb{R}_{++}$, there exists $L>0$ such that if $(p,m_1),(p,m_2)\in C$, then
\[\|f(p,m_1)-f(p,m_2)\|\le L|m_1-m_2|.\]
Moreover, we say that $f$ is {\bf locally Lipschitz} if for every compact subset $C\subset \mathbb{R}^n_{++}\times \mathbb{R}_{++}$, there exists $L>0$ such that if $(p_1,m_1),(p_2,m_2)\in C$, then
\[\|f(p,m_1)-f(p,m_2)\|\le L\|(p_1,m_1)-(p_2,m_2)\|.\]
Obviously, every locally Lipschitz CoD is income-Lipschitzian, and it is known that every continuously differentiable CoD is locally Lipschitz.

Suppose that $f$ is a CoD that is differentiable at $(p,m)$. Define
\[s_{ij}(p,m)=\frac{\partial f_i}{\partial p_j}(p,m)+\frac{\partial f_i}{\partial m}(p,m)f_j(p,m),\]
and let $S_f(p,m)$ denote the $n\times n$ matrix whose $(i,j)$-th component is $s_{ij}(p,m)$. This matrix-valued function $S_f(p,m)$ is called the {\bf Slutsky matrix}. An alternative expression of this matrix is as follows:
\[S_f(p,m)=D_pf(p,m)+D_mf(p,m)f^T(p,m),\]
where $f^T(p,m)$ denotes the transpose of $f(p,m)$. If $f$ is locally Lipschitz, then by Rademacher's theorem, $f$ is differentiable almost everywhere, and thus, the Slutsky matrix is defined almost everywhere. We say that $f$ satisfies (S) if $S_f(p,m)$ is symmetric almost everywhere, and satisfies (NSD) if $S_f(p,m)$ is negative semi-definite almost everywhere.

For a CoD $f$, we define $R(f)$ as the range of $f$. That is,
\[R(f)=\{x\in \Omega|x=f(p,m)\mbox{ for some }(p,m)\in \mathbb{R}^n_{++}\times \mathbb{R}_{++}\}.\]

Finally, suppose that $f$ is a CoD such that $R(f)$ includes $\mathbb{R}^n_{++}$. Define
\[G^f(x)=\left\{p\in \mathbb{R}^n_{++}\left|\sum_ip_i=1,\ f(p,p\cdot x)=x\right.\right\}\]
for each $x\in \mathbb{R}^n_{++}$. This multi-valued function is called the {\bf inverse demand correspondence} of $f$. We say that $f$ satisfies the {\bf C axiom} if $G^f$ is compact-valued, convex-valued, and upper hemi-continuous on $\mathbb{R}^n_{++}$.

\section{Preliminary Result: Constructing a Reverse Calculation Method}
We first construct a rigorous and effective method for calculating a utility function that corresponds to the given CoD.

\vspace{12pt}
\noindent
{\bf Proposition 1}. Suppose that $f$ is a locally Lipschitz CoD that satisfies Walras' law, (S), and (NSD). Fix $\bar{p}\gg 0$, and define $u_{f,\bar{p}}(x)$ as follows. First, if $x\notin R(f)$, then define $u_{f,\bar{p}}(x)=0$. Second, if $x=f(p,m)$ for some $(p,m)$, then consider the following differential equation
\begin{equation}\label{eq3.1}
\dot{c}(t)=f((1-t)p+t\bar{p},c(t))\cdot (\bar{p}-p),\ c(0)=m,
\end{equation}
and define $u_{f,\bar{p}}(x)=c(1)$. Then, the following hold.
\begin{enumerate}[1.]
\item $u_{f,\bar{p}}$ is well-defined,\footnote{That is, there uniquely exists a solution $c(t)$ to (1) whose domain is $[0,1]$, and $c(1)$ is independent of the choice of $(p,m)\in f^{-1}(x)$.} and $f=f^{u_{f,\bar{p}}}$.

\item $u_{f,\bar{p}}$ is upper semi-continuous on $R(f)$.

\item If $f=f^{\succsim}$ for some weak order $\succsim$ on $\Omega$ that is upper semi-continuous on $R(f)$, then for every $x,y\in R(f)$,
\[x\succsim y\Leftrightarrow u_{f,\bar{p}}(x)\ge u_{f,\bar{p}}(y).\]
\end{enumerate}

As a corollary, we obtain the following result.

\vspace{12pt}
\noindent
{\bf Corollary 1}. Suppose that $f$ is a locally Lipschitz CoD that satisfies Walras' law. Then, the following four statements are equivalent.
\begin{enumerate}[(i)]
\item $f=f^{\succsim}$ for some weak order $\succsim$ on $\Omega$.

\item $f=f^{u_{f,\bar{p}}}$, where $u_{f,\bar{p}}$ is defined in Theorem 1.

\item $f$ satisfies (S) and (NSD).

\item For every $(p,m)\in \mathbb{R}^n_{++}\times\mathbb{R}_{++}$, the partial differential equation
\begin{equation}\label{eq3.2}
\nabla E(q)=f(q,E(q)),\ E(p)=m,
\end{equation}
has a unique concave solution defined on $\mathbb{R}^n_{++}$.
\end{enumerate}

\vspace{12pt}
We present a few remarks on Proposition 1 and Corollary 1. In Hosoya (2017), the same result as Proposition 1 was obtained for continuously differentiable CoDs. Hosoya (2018) showed the same result for differentiable and locally Lipschitz CoDs. Because every continuously differentiable CoD is locally Lipschitz, the latter result is a pure extension of the former. In these previous theorems, the Slutsky matrix was assumed to be symmetric and negative semi-definite at {\bf every} $(p,m)$. In contrast, our Proposition 1 only requires the symmetry and negative semi-definiteness of the Slutsky matrix at {\bf almost every} $(p,m)$. Hence, Proposition 1 is a further pure extension of Hosoya's (2018) result. However, this weakening of the assumption significantly increases the difficulty of the proof for the following two reasons. First, the term $f((1-t)p+t\bar{p},c(t))$ appears in (\ref{eq3.1}). However, the set $A=\{((1-t)p+t\bar{p},c(t))|t\in [0,1]\}$ is a null set with respect to the Lebesgue measure. Hence, the Slutsky matrix may be undefined at all points of $A$. This fact renders many techniques used in related research inapplicable. Second, the classical techniques that derive such a result were constructed by Hurwicz and Uzawa (1971). However, in Hurwicz--Uzawa's proof, the weak axiom of revealed preference was first derived (Lemma 5 in their paper), and then the main result was proved using the weak axiom. Because the claim of the weak axiom includes a strict inequality, this property vanishes under limit manipulation. This indicates that the usual approximation approach would not work in the proof of Proposition 1.

In Hosoya (2021), a similar result was obtained using several techniques based on partial differential equations. The proof in Hosoya (2021) solved the above difficulties by applying perturbation techniques to a partial differential equation, although this is difficult to understand. In contrast, we construct the proof of Proposition 1 based on knowledge of ordinary differential equations.

Statement (iv) of Corollary 1 is a new necessary and sufficient condition for a CoD to be a demand function of some weak order. The strong axiom of revealed preference is necessary and sufficient for a CoD to be a demand function (Richter 1966; Mas-Colell et al. 1995), and recently it was shown that (S) and (NSD) are necessary and sufficient for a continuously differentiable CoD satisfying Walras' law to be a demand function (Hosoya, 2017). Our condition (iv) is a new alternative necessary and sufficient condition for a locally Lipschitz CoD satisfying Walras' law to be a demand function. Later, we discuss why this condition is crucial for the results presented in this paper.

Equation (\ref{eq3.2}) is deeply related to the {\bf expenditure function}. For a given weak order $\succsim$ on $\Omega$ and $x\in \Omega$, define
\[E^x(p)=\inf\{p\cdot y|y\succsim x\}.\]
This function is called the expenditure function. Indeed, $E^x$ coincides with the value function of the following minimization problem:
\begin{align*}
\min~~~~~&~p\cdot y\\
\mbox{subject to }&~y\in \Omega,\\
&~y\succsim x.
\end{align*}
This is traditionally called the {\bf expenditure minimization problem} in consumer theory. The expenditure function is concave and continuous. Moreover, if $f=f^{\succsim}$ and $f$ is continuous, and if $x=f(p,m)$, then $q\mapsto E^x(q)$ satisfies (\ref{eq3.2}). This result is usually called {\bf Shephard's lemma} (see Lemma 1 of Hosoya (2020)). Therefore, condition (i) implies condition (iv). It is obvious that condition (ii) implies condition (i), and it is easy to show that condition (iv) implies condition (iii). Finally, Proposition 1 claims that condition (iii) implies condition (ii). This is the background logic to Corollary 1.

If $f=f^{\succsim}$ and $f$ is continuous and income-Lipschitzian, then we can easily show that $p\mapsto E^x(p)$ is the unique solution to (\ref{eq3.2}), and $u_{f,\bar{p}}(x)=E^x(\bar{p})$ for all $x\in R(f)$. If $R(f)$ includes $\mathbb{R}^n_{++}$ and is open in $\mathbb{R}^n_+$, then by applying a similar proof as that of Theorem 1 in Hosoya (2020), we obtain the following result.

\vspace{12pt}
\noindent
{\bf Corollary 2}. Suppose that $f$ is a locally Lipschitz CoD that satisfies Walras' law, (S), and (NSD). Moreover, suppose that $R(f)$ includes $\mathbb{R}^n_{++}$ and is relatively open in $\Omega$. Define
\begin{equation}\label{eq3.3}
v_{f,\bar{p}}(x)=\begin{cases}
u_{f,\bar{p}}(x) & \mbox{if }x\in \mathbb{R}^n_{++},\\
\inf_{\varepsilon>0}\sup\{u_{f,\bar{p}}(y)|y\in\mathbb{R}^n_{++},\ \|y-x\|<\varepsilon\} & \mbox{if }x\notin \mathbb{R}^n_{++}.
\end{cases}
\end{equation}
Then, $f=f^{v_{f,\bar{p}}}$ and $v_{f,\bar{p}}$ is upper semi-continuous. Moreover, the following hold.
\begin{enumerate}[1)]
\item The function $f$ satisfies the C axiom if and only if $v_{f,\bar{p}}$ is continuous on $\mathbb{R}^n_{++}$.

\item If $f=f^{\succsim}$ for some upper semi-continuous weak order $\succsim$, then $v_{f,\bar{p}}$ represents $\succsim$. In particular, such a $\succsim$ must be unique.

\item $f=f^{\succsim}$ for some continuous weak order $\succsim$ if and only if $v_{f,\bar{p}}$ is continuous.
\end{enumerate}

\vspace{12pt}
\noindent
{\bf Example 1}. Consider the following CoD:
\[f(p,m)=\begin{cases}
\left(\frac{m}{p_1},0\right) & \mbox{if }p_2^2\ge 4p_1m,\\
\left(\frac{p_2^2}{4p_1^2},\frac{4p_1m-p_2^2}{4p_1p_2}\right) & \mbox{otherwise.}
\end{cases}\]
This function satisfies all requirements of Proposition 1 but is not continuously differentiable. Moreover, $R(f)=\{(x_1,x_2)\in \mathbb{R}^2_+|x_1>0\}$, and so this function also satisfies all requirements of Corollary 2. Set $\bar{p}=(1,1)$, and choose any $x\in R(f)$. Then, $x=f(p,m)$ for some $(p,m)\in \mathbb{R}^2_{++}\times \mathbb{R}_{++}$. If necessary, we can replace $p_2$ with $\min\{2\sqrt{p_1m},p_2\}$, and thus we can assume that $p_2^2\le 4p_1m$. Moreover, again if necessary, we can replace $(p,m)$ with $\frac{1}{p_2}(p,m)$, and thus we can assume $p_2=1$ and $4p_1m\ge 1$. Let us try to solve (\ref{eq3.1}) and determine $u_{f,\bar{p}}(x)$ and $v_{f,\bar{p}}(x)$.

First, define
\[f^1(q,w)=\left(\frac{w}{q_1},0\right),\ f^2(q,w)=\left(\frac{q_2^2}{4q_1^2},\frac{4q_1w-q_2^2}{4q_1q_2}\right),\]
and consider
\begin{equation}\label{eq3.4}
\dot{c}_i(t)=f^i((1-t)p+t\bar{p},c_i(t))\cdot (\bar{p}-p).
\end{equation}
To solve (\ref{eq3.4}), we have that
\[c_1(t)=c_1(s)\frac{p_1+t(1-p_1)}{p_1+s(1-p_1)},\]
and because $p_2=\bar{p}_2=1$,
\[c_2(t)=c_2(s)-\frac{1}{4}\left[\frac{1}{p_1+t(1-p_1)}-\frac{1}{p_1+s(1-p_1)}\right].\]
In particular, if $s=0$ and $c_2(0)=m$, then
\[c_2(t)=m-\frac{1}{4}\left[\frac{1}{p_1+t(1-p_1)}-\frac{1}{p_1}\right].\]

Second, suppose that $4c_2(1)\ge 1$, where $c_2(0)=m$. By our initial assumption, $4p_1c_2(0)\ge 1$. Moreover, $(p_1+t(1-p_1))c_2(t)$ is monotone in $t$.\footnote{Recall that $(p_1+t(1-p_1))c_2(t)=(p_1+t(1-p_1))(m+1/4p_1)-1/4$.} Therefore, $c(t)=c_2(t)$ is a solution to (\ref{eq3.1}) defined on $[0,1]$, and hence
\[c(1)=c_2(1)=m+\frac{1-p_1}{4p_1}.\]
The condition $4c_2(1)=4m+\frac{1-p_1}{p_1}\ge 1$ is equivalent to
\begin{equation}\label{eq3.5}
4p_1m+1\ge 2p_1.
\end{equation}
Because
\[x_1=\frac{1}{4p_1^2},\ x_2=m-\frac{1}{4p_1},\]
we find that (\ref{eq3.5}) is equivalent to
\[\sqrt{x_1}+x_2\ge \frac{1}{2}.\]
Moreover,
\[c(1)=\sqrt{x_1}+x_2-\frac{1}{4}.\]
Therefore, we obtain
\[u_{f,\bar{p}}(x)=\sqrt{x_1}+x_2-\frac{1}{4},\]
if $x_1>0$ and $\sqrt{x_1}+x_2\ge \frac{1}{2}$.

Third, suppose that $4c_2(1)<1$, where $c_2(0)=m$. By the same argument as above, this assumption is equivalent to
\[\sqrt{x_1}+x_2<\frac{1}{2}.\]
If $1\ge p_1$, then $(p_1+t(1-p_1))c_2(t)$ is either increasing or constant, and hence $4p_1m<1$, which contradicts our initial assumption. Thus, we have that $1<p_1$. We guess that $c(t)=c_2(t)$ on $[0,t^*]$ and $c(t)=c_1(t)$ on $[t^*,1]$, where $c(t^*)=c_1(t^*)=c_2(t^*)$ and $\dot{c}_1(t^*)=\dot{c}_2(t^*)$. Then,
\[\frac{c(t^*)(1-p_1)}{p_1+t^*(1-p_1)}=\dot{c}_1(t^*)=\dot{c}_2(t^*)=\frac{1-p_1}{4(p_1+t^*(1-p_1))^2},\]
and thus,
\[c(t^*)=c_1(t^*)=c_2(t^*)=\frac{1}{4(p_1+t^*(1-p_1))}.\]
Then, 
\[c_2(t^*)=m-\frac{1}{4}\left[\frac{1}{p_1+t^*(1-p_1)}-\frac{1}{p_1}\right]=\frac{1}{4(p_1+t^*(1-p_1))},\]
and hence, we obtain
\[t^*=\frac{p_1}{1-p_1}\left[\frac{1-4p_1m}{4p_1m+1}\right].\]
Because $1-p_1<0$ and $1\le 4p_1m$, we have that $t^*\ge 0$. Moreover, because $(p_1+t(1-p_1))c_2(t)$ is decreasing on $[0,1]$, $4c_2(1)<1$, and $4(p_1+t^*(1-p_1))c_2(t^*)=1$, we have that $t^*<1$. Therefore, $t^*\in [0,1]$. Hence,
\[c(t)=\begin{cases}
m-\frac{1}{4}\left[\frac{1}{p_1+t(1-p_1)}-\frac{1}{p_1}\right] & \mbox{if }t\le t^*,\\
\frac{(p_1+t(1-p_1))}{4(p_1+t^*(1-p_1))^2} & \mbox{if }t\ge t^*.\\
\end{cases}\]
We can check that this $c(t)$ is actually the solution. In particular,
\[c(1)=\frac{1}{4(p_1+t^*(1-p_1))^2}=\frac{16p_1^2m^2+8p_1m+1}{16p_1^2}.\]
Because
\[x_1=\frac{1}{4p_1^2},\ x_2=m-\frac{1}{4p_1},\]
we have that
\[c(1)=(\sqrt{x_1}+x_2)^2=(\sqrt{x_1}+x_2)^2.\]
Therefore, if $x_1>0$ and $\sqrt{x_1}+x_2<\frac{1}{2}$, then
\[u_{f,\bar{p}}(x)=(\sqrt{x_1}+x_2)^2.\]
In conclusion, we obtain
\[u_{f,\bar{p}}(x)=\begin{cases}
\sqrt{x_1}+x_2-\frac{1}{4} & \mbox{if }x_1>0,\ \sqrt{x_1}+x_2\ge \frac{1}{2},\\
(\sqrt{x_1}+x_2)^2 & \mbox{if }x_1>0,\ \sqrt{x_1}+x_2<\frac{1}{2},\\
0 & \mbox{if }x_1=0.
\end{cases}\]
Of course, in this case,
\[v_{f,\bar{p}}(x)=\begin{cases}
\sqrt{x_1}+x_2-\frac{1}{4} & \mbox{if }\sqrt{x_1}+x_2\ge \frac{1}{2},\\
(\sqrt{x_1}+x_2)^2 & \mbox{if }\sqrt{x_1}+x_2<\frac{1}{2}.
\end{cases}\]
This completes our calculation.

\section{Main Result: Continuity of Calculation}
We are now able to calculate a utility function $u_{f,\bar{p}}$ from a CoD $f$. However, in the real world, we can only obtain finite data about $f$, and because $f$ includes infinite data, we cannot determine $f$ rigorously. Hence, our CoD $f$ must be considered as an estimated value of the true demand function. In this view, we need a continuity result: that is, we require that if $f'$ is near to $f$, then $u_{f',\bar{p}}$ is also near to $u_{f,\bar{p}}$. If this condition is violated, then we are confronted with a methodological difficulty---ensuring ``consistency'' for the estimated utility function becomes hard.

We should explain this point in detail. Consider some estimation problem with the true value $x\in X$. Suppose that there is a given estimation method, and for some data with size $N$, let $x_N$ be the estimated value of $x$. Then, $x_N$ is a random variable on $X$. This estimation method is said to be {\bf consistent} if $x_N$ converges to $x$ in probability as $N\to \infty$.

Suppose that there is an estimation method for the true demand function $f$, and $f_N$ is an estimated value of $f$ for some data set of size $N$. Suppose also that this estimation method is consistent with respect to some topology on the space of demand functions. For each $f_N$, we can calculate the utility function $u_{f_N,\bar{p}}$, and thus, $u_{f_N,\bar{p}}$ can be treated as an estimated value of the ``true utility function'' $u_{f,\bar{p}}$. Our question is as follows: does $u_{f_N,\bar{p}}$ converge to $u_{f,\bar{p}}$? If not, our estimation method violates the consistency condition, and is thus not useful.

Hence, the continuity of $u_{f,\bar{p}}$ with respect to $f$ is very important. In this regard, we first show the following result.

\vspace{12pt}
\noindent
{\bf Theorem 1}. Suppose that $(f^k)$ is a sequence of locally Lipschitz demand functions that satisfy Walras' law, and for every compact set $C\subset \mathbb{R}^n_{++}\times\mathbb{R}_{++}$, $f^k$ converges to a CoD $f$ uniformly on $C$ as $k\to \infty$. If $f$ is locally Lipschitz, then $f$ is also a demand function.\footnote{It is obvious that, under the assumption of this theorem, $f$ satisfies Walras' law.}

\vspace{12pt}
As a corollary, we obtain an important result. Let $\Delta_{\nu}=[\nu^{-1},\nu]^{n+1}$. Define a metric
\[\rho(f,f')=\sum_{\nu=1}^{\infty}\frac{1}{2^{\nu}}\min\left\{\sup_{(p,m)\in \Delta_{\nu}}\|f(p,m)-f'(p,m)\|,1\right\},\]
where $f,f'$ are CoDs. We can easily show that $\rho$ is a metric in the space of CoDs, and a sequence $(f^k)$ converges to $f$ with respect to $\rho$ if and only if, for every compact set $C\subset \mathbb{R}^n_{++}\times \mathbb{R}_{++}$, $(f^k)$ converges to $f$ uniformly on $C$.

Suppose that $L=(L_{\nu})$ is a sequence of positive real numbers. Define $\mathscr{F}_L$ as the set of demand functions $f$ that satisfy Walras' law and the following inequality:
\[\|f(p,m)-f(q,w)\|\le L_{\nu}\|(p,m)-(q,w)\|\]
for all $\nu\in \mathbb{N}$ and $(p,m),(q,w)\in \Delta_{\nu}$.

\vspace{12pt}
\noindent
{\bf Corollary 3}. The space $\mathscr{F}_L$ is compact with respect to the metric $\rho$.

\vspace{12pt}
To ensure the uniqueness of the upper semi-continuous weak order corresponding to $f$, we need to use Corollary 2. Thus, an additional assumption is needed: that is, $R(f)$ must include $\mathbb{R}^n_{++}$ and be open in $\mathbb{R}^n_+$. Suppose that $(f^k)$ is a sequence on $\mathscr{F}_L$ that converges to $f$, and for every $k$, $f^k$ satisfies all requirements in Corollary 2. Does $f$ also satisfy the requirements of Corollary 2? Unfortunately, the following example indicates that the answer is negative.

\vspace{12pt}
\noindent
{\bf Example 2}. Consider the class of CES utility functions:
\[u^{\sigma}(x)=(x_1^{\sigma}+x_2^{\sigma})^{\frac{1}{\sigma}},\]
where $\sigma<1$ and $\sigma\neq 0$. The corresponding demand function is
\[f_i^{\sigma}(p,m)=\frac{p_i^{\frac{-1}{1-\sigma}}m}{p_1^{\frac{-\sigma}{1-\sigma}}+p_2^{\frac{-\sigma}{1-\sigma}}}.\]
We assume that $\sigma<0$. To differentiate this function, for $j\neq i$ and $(p,m)\in \Delta_{\nu}$, we have that
\begin{align*}
\left|\frac{\partial f_i^{\sigma}}{\partial p_i}(p,m)\right|=&~\frac{[(1-\sigma)p_i^{\frac{-2}{1-\sigma}}+p_i^{\frac{-2+\sigma}{1-\sigma}}p_j^{\frac{-\sigma}{1-\sigma}}]m}{(1-\sigma)(p_1^{\frac{-\sigma}{1-\sigma}}+p_2^{\frac{-\sigma}{1-\sigma}})^2}\le \frac{\nu^5}{2},\\
\left|\frac{\partial f_i^{\sigma}}{\partial p_j}(p,m)\right|=&~\frac{-\sigma p_i^{\frac{-1}{1-\sigma}}p_j^{\frac{-1}{1-\sigma}}m}{(1-\sigma)(p_1^{\frac{-\sigma}{1-\sigma}}+p_2^{\frac{-\sigma}{1-\sigma}})^2}\le\frac{\nu^5}{4},\\
\left|\frac{\partial f_i^{\sigma}}{\partial m}(p,m)\right|=&~\frac{p_i^{\frac{-1}{1-\sigma}}}{p_1^{\frac{-\sigma}{1-\sigma}}+p_2^{\frac{-\sigma}{1-\sigma}}}\le\frac{\nu^2}{2}.
\end{align*}
Thus, if we define $L_{\nu}=\nu^5$, then $f_i^{\sigma}\in \mathscr{F}_L$. Moreover, for every $\sigma<0$, we have that $R(f^{\sigma})=\mathbb{R}^2_{++}$. However, $f^{\sigma}$ converges to a function $f$ as $\sigma\to -\infty$ with respect to the metric $\rho$, where
\[f_1(p,m)=f_2(p,m)=\frac{m}{p_1+p_2},\]
and $R(f)=\{(c,c)|c>0\}$. This fact implies that the limit manipulation in $\mathscr{F}_L$ may shrink the range of the demand function.

\vspace{12pt}
The above example shows that, for our purpose, an additional assumption is needed. One of the easiest ways to solve this problem is to assume that the limit CoD $f$ satisfies all assumptions of Corollary 2. The result is as follows.

\vspace{12pt}
\noindent
{\bf Theorem 2}. Let $(f^k)$ be a sequence of locally Lipschitz demand functions such that $f^k$ satisfies Walras' law for all $k$ and it converges to a locally Lipschitz demand function $f$ with respect to $\rho$. Suppose that $R(f^k)$ includes $\mathbb{R}^n_{++}$ and $f^k$ satisfies the C axiom for all $k$, and that $f$ also satisfies these conditions. Then, for every compact set $D\subset \mathbb{R}^n_{++}$,
\[\sup_{x\in D}|u_{f^k,\bar{p}}(x)-u_{f,\bar{p}}(x)|\to 0\]
as $k\to \infty$.

\vspace{12pt}
The definition of $v_{f,\bar{p}}$ in Corollary 2 only depends on the values of $u_{f,\bar{p}}$ on $\mathbb{R}^n_{++}$. Hence, it seems to show that $v_{f^k,\bar{p}}$ converges to $v_{f,\bar{p}}$ pointwise, where $v_{f^k,\bar{p}}$ is as defined in Corollary 2. However, there are several technical difficulties that mean we cannot obtain such a result. Instead, we can show the following result.

\vspace{12pt}
\noindent
{\bf Corollary 4}. In addition to the assumptions of Theorem 2, suppose that $R(f)$ and all $R(f^k)$ are relatively open in $\Omega$. Then, $\limsup_{k\to \infty}v_{f^k,\bar{p}}(x)\le v_{f,\bar{p}}(x)$ for every $x\in \Omega$.

\vspace{12pt}
Because $v_{f^k,\bar{p}}(x)\ge 0$ for every $x\in \Omega$, if $v_{f,\bar{p}}(x)=0$, then $v_{f^k,\bar{p}}(x)$ converges to $v_{f,\bar{p}}(x)$. However, if $v_{f,\bar{p}}(x)>0$, whether $v_{f^k,\bar{p}}(x)$ converges to $v_{f,\bar{p}}(x)$ or not is unknown. Indeed, we have the following example in which $\lim_{k\to \infty}v_{f^k,\bar{p}}(x)\neq v_{f,\bar{p}}(x)$.

\vspace{12pt}
\noindent
{\bf Example 3}. Suppose that $h:\mathbb{R}_{++}\to \mathbb{R}_{++}$ is $C^{\infty}$, nondecreasing, and $\lim_{c\to 0}h(c)=0,\lim_{c\to \infty}h(c)=\infty$. Choose any $(x_1,x_2)\in \mathbb{R}^2_{++}$, and consider the equation
\[(x_1^{\frac{1}{1+\frac{1}{c}}}+x_2^{\frac{1}{1+\frac{1}{c}}})^{1+\frac{1}{c}}=h(c).\]
By the same arguments as in the proof of Example 4 in Hosoya (2020), we can show that there exists a unique solution $c^*>0$ to the above equation, and if we define $u^h(x_1,x_2)=c^*$, then $u^h$ is $C^{\infty}$, monotone, and strictly quasi-concave. If $(x_1,x_2)\in \mathbb{R}^2_+\setminus\mathbb{R}^2_{++}$, then we define
\[u^h(x_1,x_2)=\inf_{\varepsilon>0}\sup\{u^h(x_1',x_2')|(x_1',x_2')\in\mathbb{R}^2_{++},\ \|(x_1,x_2)-(x_1',x_2')\|<\varepsilon\}.\]
By the same arguments as in the proof of Corollary 2, we can show that $u^h$ is upper semi-continuous on $\mathbb{R}^2_+$. Moreover, again by the same arguments as in the proof of Example 4 in Hosoya (2020), we can show that $f^{u^h}$ is also $C^{\infty}$, and $R(f^{u^h})=\mathbb{R}^2_{++}$. Furthermore, if $h'\to h$ in the sense of $C^2$, then we can easily show that $u^{h'}\to u^h$ with respect to the local $C^2$ topology on $\mathbb{R}^n_{++}$, and by Proposition 2.7.2 of Mas-Colell (1985), we have that $f^{h'}\to f^h$ with respect to the metric $\rho$.

Let $h:\mathbb{R}_+\to \mathbb{R}_+$ be a $C^{\infty}$ nondecreasing function such that $h(0)=0, h(1)=h(2)=1$, $h'(c)>0$ if $c\notin [1,2]$, and $\lim_{c\to \infty}h(c)=+\infty$. Additionally, let $\eta:\mathbb{R}_+\to \mathbb{R}_+$ be a $C^{\infty}$ function such that $\eta(c)\equiv 0$ on $[0,1]$ and $[4,+\infty[$, $\eta(c)$ is increasing on $[1,2]$, constant on $[2,3]$, decreasing on $[3,4]$, and $\max_{c\in [1,4]}|\eta'(c)|<\min_{c\in [3,4]}h'(c)$. Define $h^k(c)=h(c)+k^{-1}\eta(c)$. Then, $h^k\to h$ with respect to the $C^2$ topology, and thus $f^{h^k}\to f^h$ with respect to $\rho$. Let $\bar{p}=(1,1)$. Then, we can easily check that
\[v_{f^{h^k},\bar{p}}(1,0)=\frac{1}{2},\]
for all $k$, and
\[v_{f^h,\bar{p}}(1,0)=\frac{1}{\sqrt{2}},\]
which implies that $\lim_{k\to \infty}v_{f^{h^k},\bar{p}}(1,0)\neq v_{f^h,\bar{p}}(1,0)$.

\vspace{12pt}
We now present another completeness result. Choose a sequence $M=(M_{\nu})$ of positive real numbers and define $\mathscr{F}_{L,M}$ as the set of all $f\in \mathscr{F}_L$ such that $R(f)$ includes $\mathbb{R}^n_{++}$, $f$ satisfies the C axiom, and if $x\in ]\nu^{-1},\nu[^n$, then for all $p\in G^f(x)$ and $i\in \{1,...,n\}$, $p_i\ge M_{\nu}$.

\vspace{12pt}
\noindent
{\bf Corollary 5}. $\mathscr{F}_{L,M}$ is compact under the metric $\rho$. Moreover, if $(f^k)$ is a sequence on $\mathscr{F}_{L,M}$ that converges to $f$ with respect to $\rho$, then for every compact set $D\subset \mathbb{R}^n_{++}$,
\[\sup_{x\in D}|u_{f^k,\bar{p}}(x)-u_{f,\bar{p}}(x)|\to 0\]
as $k\to \infty$.

\vspace{12pt}
In Theorem 2, the C axiom is required. Corollary 2 states that if $R(f)$ is relatively open in $\Omega$, this axiom is equivalent to the continuity of $v_{f,\bar{p}}$ on $\mathbb{R}^n_{++}$. Indeed, the requirement that $R(f)$ is relatively open in $\Omega$ is not used in the proof of this fact. Therefore, under the assumptions of Theorem 2, the same proof shows that $u_{f^k,\bar{p}}$ and $u_{f,\bar{p}}$ are continuous on $\mathbb{R}^n_{++}$. We use the continuity of $u_{f,\bar{p}}$ on $\mathbb{R}^n_{++}$ in the proof of Theorem 2; see Lemma 4. However, $v_{f,\bar{p}}$ is not necessarily continuous on $\Omega$ itself, even if $R(f)$ is open; see Example 4 of Hosoya (2020).

If $f$ is a demand function that is continuously differentiable on $P\equiv f^{-1}(\mathbb{R}^n_{++})$ and the rank of $S_f(p,m)$ is always $n-1$ on $P$, then we can show that the inverse demand correspondence $G^f(x)$ is a single-valued continuously differentiable function, and thus the C axiom is automatically satisfied. For a proof, see Proposition 1 of Hosoya (2013). In this connection, if $(f^k)$ is a sequence on $\mathscr{F}_L$ such that every $f^k$ is continuously differentiable and the rank of $S_{f^k}(p,m)$ is always $n-1$ on $(f^k)^{-1}(\mathbb{R}^n_{++})$, then there exists a sequence $M=(M_{\nu})$ such that every $f^k$ is in $\mathscr{F}_{L,M}$ if and only if $\inf_k\min_i\min_{x\in C}G_i^{f^k}(x)>0$ for every compact set $C\subset \mathbb{R}^n_{++}$. This is another sufficient condition for the limit function $f$ of $(f^k)$ to satisfy all requirements of Theorem 2.

Finally, we present a result for pointwise convergence. In many cases, the solution function of a differential equation does not exhibit good behavior with respect to pointwise convergence. However, in this case, the problem can be avoided using the equicontinuity of $\mathscr{F}_L$. Hence, the following theorem holds.

\vspace{12pt}
\noindent
{\bf Theorem 3}. Suppose that $(f^k)$ is a sequence in $\mathscr{F}_L$ that converges pointwise to $f$. Then, $f\in \mathscr{F}_L$. In particular, if $f^k\in \mathscr{F}_{L,M}$ for all $k$, then $f\in \mathscr{F}_{L,M}$, and for every compact set $D\subset \mathbb{R}^n_{++}$,
\[\sup_{x\in D}|u_{f^k,\bar{p}}(x)-u_{f,\bar{p}}(x)|\to 0\]
as $k\to \infty$.

\vspace{12pt}
This result is unexpected in some ways. Previous results in this context have usually required a stronger topology in the space of demand functions to prove convergence in some topology of the space of utility functions. For example, in Hosoya (2017), convergence with respect to a uniform topology in the space of utility function could only be proved if the $C^1$ topology is equipped in the space of demand functions. In Theorem 3, however, this relationship is reversed.

As a final note, we mention the closed convergence topology of weak orders. If the shapes of utility functions are specified for some set of weak orders, then in most cases, the compact convergence of the utility function is equivalent to the convergence in the closed convergence topology of the weak order. Hence, for example, it is quite easy to derive the convergence result in the closed convergence topology from Theorems 2-3. In this connection, in econometric studies that use statistical models that require a particular shape for the utility function, we can inversely derive the compact convergence of their utility function from the convergence of corresponding orders in the closed convergence topology. In this sense, the use of a specified shape of the utility function is not a disadvantage for Theorems 2-3.

\section{Discussion}
\subsection{Comparison with Related Literature}
The history of integrability theory begins with Antonelli (1886). This theory aims to calculate a utility function from the consumer's purchase behavior. Hurwicz (1971) classified this theory into two categories, indirect and direct. The indirect approach involves deriving the inverse demand function from the purchase behavior and then calculating a utility function by finding some function that satisfies Lagrange's first-order conditions for the inverse demand function. The direct approach derives the demand function from the purchase behavior, solves Shephard's lemma as a partial differential equation to calculate the expenditure function, and then calculates the utility function from this expenditure function. Antonelli (1886) used the indirect approach, as did most of the classical results (Pareto 1906; Samuelson 1950; Katzner 1970; Debreu 1972). In contrast, Hurwicz and Uzawa (1971) obtained a classical result using the direct approach. Hosoya (2013) provides an example of the indirect approach, while Hosoya (2017) is an example of the direct approach. Proposition 1 in this paper is categorized as a direct approach.

To understand the position of this paper in integrability theory using the direct approach, let us look at the classical result of Hurwicz--Uzawa. They showed that if a CoD is differentiable and locally Lipschitz, satisfies Walras' law, (S), (NSD), and a condition called the ``strong income-Lipschitzian'' requirement, then it is a demand function. Although they did not specify how to derive the utility function in their theorem, the utility function that appears in their proof essentially coincides with our $u_{f,\bar{p}}$. Following their paper, several studies attempted to remove the ``strong income-Lipschitzian'' requirement, and Hosoya (2017, 2018) finally succeeded in doing so.

Let us explain the logic that allowed us to eliminate the strong income-Lipschitzian requirement. In the proof of Hurwicz--Uzawa's Theorem 2, this condition is only used to derive the existence of the solution to the partial differential equation (\ref{eq3.2}). In fact, the necessary and sufficient condition for the existence of local solutions to (\ref{eq3.2}) is (S), which was proved in Theorem 10.9.4 of Dieudonne (1969). Hurwicz--Uzawa constructed a similar proof to that of Nikliborc (1929) to show the existence of a global solution to (\ref{eq3.2}), in which the strong income-Lipschitzian condition and (S) were used. From this and (NSD), they then proved the claim considered in Step 4 of the proof of our Proposition 1. Once the existence of global solutions to (\ref{eq3.2}) and the statement in Step 4 have been shown, we no longer require the differentiability of $f$ to prove this theorem. In contrast, Hosoya (2017) brought (NSD) to the proof of the existence of the global solution to (\ref{eq3.2}) and showed the existence of global solutions without the strong income-Lipschitzian condition. This is why we can eliminate the strong income-Lipschitzian condition from this theorem. Hosoya (2017) treated continuously differentiable CoDs, and later this result was extended to differentiable and locally Lipschitz CoDs (Hosoya, 2018).

The present paper removes even the requirement for differentiability and assumes only locally Lipschitz conditions for CoDs. First, we explain why the removal of differentiability is needed. Classical methods for estimating demand functions have been studied for a long time (for example, Deaton (1986) contains a detailed description of parameter estimation methods for demand functions). More recently, Blundell et al. (2017) presented a method for estimating the demand function without parametrization. Obviously, such estimation methods must be verified to satisfy consistency. As already mentioned, consistency means that the estimated value $x_N$ from a dataset of size $N$ converges to the true value $x$ in probability as $N\to \infty$. Therefore, it must be confirmed that the estimated value $f_N$ of the demand function converges to the true demand function $f$. However, because both $f_N$ and $f$ are functions, the convergence concept can make an important difference. As far as we know, in econometric theory, either pointwise convergence or uniform convergence on compacta is used in most research. However, the space of the differentiable demand functions is not closed in both topologies.

In this connection, we are interested in whether our constructed utility function $u_{f,\bar{p}}$ satisfies consistency. In other words, we argue whether $u_{f_N,\bar{p}}$ converges to $u_{f,\bar{p}}$ if $f_N$ converges to $f$. Hosoya (2017) showed that if $f_N$ converges to $f$ in the sense of $C^1$ on any compact set, then $u_{f_N,\bar{p}}$ converges to $u_{f,\bar{p}}$ uniformly on any compact set. However, as already mentioned, there is almost no estimation method in existing studies that treats $C^1$ convergence in the estimation of the demand function. Hence, this result is not practical. Therefore, we want to consider the case in which $f_N$ converges to $f$ uniformly on any compact set. Then, the following problem arises. Because we can choose our estimation method, we may be able to choose one that makes $f_N$ differentiable. However, $f$ is the limit of $f_N$ with respect to the uniform topology, and thus, $f$ is not necessarily differentiable. If $u_{f,\bar{p}}$ can only be defined on differentiable functions, then, for nondifferentiable $f$, $u_{f,\bar{p}}$ cannot be defined in the first place, and the convergence problem of $u_{f_N,\bar{p}}$ becomes nonsensical.

Proposition 1 in this paper fundamentally resolves this problem. If we construct an estimation method so that the estimated value $f_N$ is included in $\mathscr{F}_L$, then the limit $f$ is automatically a locally Lipschitz function. Thus, Proposition 1 can be applied, and we can define $u_{f,\bar{p}}$. Furthermore, we have already confirmed in Theorems 1 and 2 that the space of the demand functions we wish to use, such as $\mathscr{F}_L$ and $\mathscr{F}_{L,M}$, is closed with respect to uniform convergence,\footnote{Actually, they are not only closed, but also compact. See Corollary 3.} and thus, the possibility that the true value is an inconvenient function can be eliminated at the construction stage of the estimation method. Theorem 3 proves that the same holds for pointwise convergence.

These arguments only make sense if Proposition 1 can be verified. Therefore, Proposition 1 is the crux of this study. However, the removal of differentiability poses two difficulties in the proof of Proposition 1. First, as already mentioned, the Lebesgue measure of the trajectory of $((1-t)p+t\bar{p},c(t))$ in (\ref{eq3.1}) is $0$. By Rademacher's theorem, any locally Lipschitz function is differentiable almost everywhere. However, $f$ may be nondifferentiable at every point on the above trajectory, because this trajectory is a null set. Even if it is differentiable, the Slutsky matrix may not have good properties. This means that most of the tools used in integrability theory up to now cannot be used as they are. This problem is solved by perturbing the solution to the differential equation by the income (see Lemma 2 in the proof). Another difficulty arises in Step 4 of the proof of Proposition 1. This step claims that $p\cdot y>m$. However, this inequality is strong, and the simple perturbation technique can no longer be used in the proof, because the strong inequality is replaced by a weak inequality through limit manipulation. This can be solved by rigorous evaluation of the inequality of the perturbed trajectory, but this evaluation is not straightforward (see the proof of Step 4).

Corollary 1 is derived from Proposition 1. This is one of the newest results in integrability theory. As implied by Houthakker (1950) and shown by Uzawa (1960) and Richter (1966), the strong axiom of revealed preference is a necessary and sufficient condition for a CoD to be a demand function. For this result, no topological condition is imposed on the CoD. If a CoD satisfies Walras' law and is continuously differentiable, then (S)+(NSD) is a necessary and sufficient condition for it to be a demand function, as shown by Hosoya (2017). Corollary 1 demonstrates that the same result holds when the CoD is not differentiable, but only locally Lipschitz. The most important thing about Corollary 1, however, is that it presents another necessary and sufficient condition for a CoD to be a demand function, namely the existence of a concave global solution to the partial differential equation (\ref{eq3.2}).

To illustrate the importance of this result, we begin by recalling the strong axiom of revealed preference. A CoD $f$ satisfies the strong axiom of revealed preference if and only if, for every finite sequence $x^1,...,x^{\ell}$ such that $x^i=f(p^i,m^i)$ and $p^i\cdot x^{i+1}\le m^i$, $p^{\ell}\cdot x^1>m^{\ell}$. The problem is that there is a strong inequality in this claim. Even if $f^k$ satisfies this condition, this strong inequality changes to a weak inequality in the limit $f$, and thus, we cannot determine whether $f$ is actually a demand function. This implies that the strong axiom of revealed preference cannot be used to prove results such as Theorem 1.

A similar problem arises when we discuss this problem using conditions (S) and (NSD). Even if $f^k$ converges to $f$ with respect to the metric $\rho$, it is uncertain whether the derivatives converge. Therefore, even if $f^k$ satisfies (S) and (NSD), $f$ may violate (S) or (NSD), and so we cannot prove that $f$ is a demand function. If we change the metric and use a stronger topology, we can show that (S) and (NSD) hold in the limit $f$. In this case, however, it becomes difficult to find results in econometric theory corresponding to such a topology. Therefore, these conditions are also undesirable.

Condition (iv) of Corollary 1 fundamentally resolves this problem. Indeed, in the proof of Theorem 1, we confirm that $f$ satisfies condition (iv). This property is not broken by convergence with respect to $\rho$, which makes such a proof possible. Hence, the remainder of our results depends on condition (iv).

Theorems 2 and 3 require the C axiom. This axiom was first discovered by Hosoya (2017), and was therefore not used by Hosoya (2015) to show a result similar to Theorem 2 in integrability theory using the indirect approach. Because $G^f(x)$ is assumed to be a single-valued, continuously differentiable function in the indirect approach, this axiom automatically holds. This is why the C axiom does not appear in Hosoya (2015). The C axiom is known to be equivalent to another axiom called the NLL axiom. In Theorem 2 of Hosoya (2020), it was shown that, for an income-Lipschitzian demand function $f$ that satisfies Walras' law, $f=f^u$ for some function $u:\Omega\to \mathbb{R}$ such that $u$ is continuous on $\mathbb{R}^n_{++}$ if and only if $f$ satisfies the C axiom. In this paper, this result is required to prove Theorem 2 (see Lemma 4).

Finally, we make an important statement. Research on estimation methods for demand functions can be separated into two types. The first type specifies the shape of the corresponding utility function, whereas the second type does not specify any particular shape. Of the research already mentioned, Deaton (1986) does not specify the utility function, but Blundell et al. (2017) do to some extent. In this connection, some readers may think that this study is not useful in research that specifies the shape of the utility function. However, this is not the case because, even if the shape of the utility function is assumed, the estimate of the utility function associated with the demand function must represent the same order as our utility function $u_{f,\bar{p}}$. Hence, by a consistency result for $u_{f,\bar{p}}$, we can almost automatically obtain a consistency result for their utility function by the method in the last paragraph of the previous section.

\subsection{Several Open Problems}
In this paper, we have attempted to produce the desired results as far as possible. However, there remain several problems that we cannot solve. Here, we describe a few of them that we consider important.

First, in Corollary 1, we proved the equivalence of conditions (i) and (iv) by assuming that $f$ is locally Lipschitz. Can this equivalence also be proved when $f$ is continuous and income-Lipschitzian? If so, then the income-Lipschitzian requirement would be sufficient for the proof of Theorem 1, which would mean that the result of Theorem 1 could be discussed on a wider space than that of locally Lipschitz CoDs. This would also strengthen Corollary 3 and Theorem 2.

Second, the question remains as to whether Corollary 4 can be strengthened. We have only shown that $\limsup_{k\to \infty}v_{f^k,\bar{p}}(x)\le v_{f,\bar{p}}(x)$, and found an example such that the inequality becomes strong. However, this inequality could perhaps be modified to an equality under some weak additional assumptions. In particular, equality may be guaranteed when $x$ is an element of $R(f)$, or $v_{f,\bar{p}}$ is continuous at $x$. If we could prove this, our result would be much better.

The third problem concerns whether the condition that $R(f)$ is an open set is necessary in the first place. In the proof of Corollary 2, we show that $v_{f,\bar{p}}$ coincides with the utility function defined in Hosoya (2020). For some technical reasons, it is necessary that $R(f)$ is an open set to guarantee that $f=f^{v_{f,\bar{p}}}$. However, there is no known counterexample of $f$ such that $R(f)$ is not an open set and $f\neq f^{v_{f,\bar{p}}}$. Perhaps $f=f^{v_{f,\bar{p}}}$ holds even when $R(f)$ is not an open set.

Fourth, there remains the task of identifying the conditions for $v_{f,\bar{p}}$ to be continuous. Condition a. of Theorem 6 in Hurwicz and Uzawa (1971) is frequently used in this context. This condition states that if $p\ge 0, p\neq 0$ and $p_i=0$ for some $i$, then for any convergent sequence $(p^k)$ to $p$ on $\mathbb{R}^n_{++}$ and any $(q,w)\in \mathbb{R}^n_{++}$, $f(p^k,v_{f,p^k}(f(q,w)))$ is unbounded. However, when discussing this condition, Hurwicz--Uzawa restrict the domain of the utility function to $R(f)$. Therefore, whether the continuity of $v_{f,\bar{p}}$ outside $R(f)$ can be guaranteed by this condition remains an open question.

Finally, although our paper only considers the classical consumer theory, there are several new consumer theories treating nonlinear or stochastic budget inequality. See, for example, Shiozawa (2016) for the former, and Allen et al. (2023) for the latter. Our study does not provide a solution to the estimation problem in those theories, and it is a future task.

\section{Conclusion}
In this study, we obtained a procedure for calculating a utility function from a given locally Lipschitz CoD that satisfies Walras' law. Using this procedure, we found two necessary and sufficient conditions for a locally Lipschitz CoD that satisfies Walras' law to be a demand function. Moreover, under the assumption that the range of this CoD includes the positive orthant and is open in the consumption space, we obtained the uniqueness result for the corresponding upper semi-continuous weak order to this CoD, and derived an upper semi-continuous utility function that represents this weak order.

Using these results, we proved a completeness result for the space of demand functions. That is, we showed that if every sequence of demand functions that is locally Lipschitz and satisfies Walras' law converges to some function with respect to the topology of compact convergence, then the limit function is also a demand function. From this result, we showed that the space of demand functions that has a uniform Lipschitz constant on any compact set is compact under this topology. 

Furthermore, we showed that if every function has a sufficiently wide range and satisfies the C axiom, then our derived utility function is continuous with respect to the demand function. Using this result, we showed that the space of demand functions that has uniform local Lipschitz constants and uniformly satisfies the C axiom is compact, and the mapping from the space of demand functions into the space of utility function is continuous. We demonstrated that a similar result holds even when we use the pointwise topology.

We also provided three examples. The first demonstrated that our calculation procedure for the utility function works well. The second example showed that the range of the CoD may shrink under limit manipulation. The third example demonstrated that our continuity result may not hold in the corner of the consumption space. We think that all examples are meaningful in this context.

Although there are many open problems concerning this area of research, we believe that the results in this paper are sufficiently strong and worthwhile for applied economic research. In particular, we think that our results provide a foundation for applying integrability theory in the field of econometric theory.

\section{Proofs}
\subsection{Mathematical Knowledge on Lipschitz Analysis and Differential Equations}
We repeatedly use Lipschitz analysis in the proofs of our theorems. However, the Lipschitz property of the solution function (defined later) for differential equations is not well known. Thus, we introduce several important properties in this subsection.

First, recall the definition of a locally Lipschitz function. Let $f:U\to \mathbb{R}^N$ be some function, where $U\subset \mathbb{R}^M$ is open. This function is said to be {\bf locally Lipschitz} if, for every compact set $C\subset U$, there exists $L>0$ such that for every $x,y\in C$,
\[\|f(x)-f(y)\|\le L\|x-y\|.\]
Because the following property is important, we present a proof in this subsection.

\vspace{12pt}
\noindent
{\bf Fact 1}. Let $f:U\to \mathbb{R}^N$, where $U\subset \mathbb{R}^M$ is open. Then, $f$ is locally Lipschitz if and only if, for every $x\in U$, there exists $r>0$ and $L>0$ such that if $y,z\in U,\ \|y-x\|\le r,$ and $\|z-x\|\le r$, then
\[\|f(y)-f(z)\|\le L\|y-z\|.\]

\vspace{12pt}
\noindent
{\bf Proof of Fact 1}. Suppose that $f$ is locally Lipschitz. For each $x\in U$, there exists $r>0$ such that $\bar{B}_r(x)\equiv \{y\in \mathbb{R}^M|\|y-x\|\le r\}\subset U$, and $\bar{B}_r(x)$ is compact. This implies that there exists $L>0$ such that if $y,z\in \bar{B}_r(x)$, then
\[\|f(y)-f(z)\|\le L\|y-z\|.\]
To prove the converse relationship, we use proof by contraposition. Suppose that $f$ is not locally Lipschitz. Then, there exist a compact set $C$ and sequences $(x^k),(y^k)$ on $C$ such that, for all $k$,
\[\|f(x^k)-f(y^k)\|>k\|x^k-y^k\|.\]
Because $C$ is compact, we can assume that $x^k\to x^*,\ y^k\to y^*$ as $k\to \infty$. Then,
\[\|f(x^*)-f(y^*)\|\ge k\|x^*-y^*\|\]
for all $k$, which implies that $x^*=y^*$. Choose any $r>0$ and $L>0$. Then, there exists $k>L$ such that $x^k,y^k\in \bar{B}_r(x^*)$, and thus the latter claim of this fact is violated. This completes the proof. $\blacksquare$

\vspace{12pt}
From the above fact, we have that every continuously differentiable function is locally Lipschitz. Of course, the converse is not true: consider $f(x)=|x|$.

The next fact is known as {\bf Rademacher's theorem}. Because the proof of this fact is long, it is omitted here.\footnote{See, for example, Heinonen (2004).}

\vspace{12pt}
\noindent
{\bf Fact 2}. Suppose that $f:U\to \mathbb{R}^N$, where $U\subset \mathbb{R}^M$ is open. If $f$ is locally Lipschitz, then it is differentiable almost everywhere.

\vspace{12pt}
Next, we explain some knowledge of ordinary differential equations (ODEs). First, consider the following ODE:
\begin{equation}\label{eq6.00}
\dot{x}(t)=g(t,x(t)),\ x(t_0)=x^*,
\end{equation}
where $g:U\to \mathbb{R}^N$ and $U\subset \mathbb{R}\times\mathbb{R}^N$ is open. We call a subset $I$ of $\mathbb{R}$ an {\bf interval} if it is a convex set containing at least two points. We say that a function $x:I\to \mathbb{R}^N$ is a {\bf solution} to (\ref{eq6.00}) if and only if 1) $I$ is an interval containing $t_0$, 2) $x(t_0)=x^*$, 3) $x$ is absolutely continuous on any compact interval $C\subset I$,\footnote{Recall that, for a function $f:[a,b]\to \mathbb{R}^N$, it is said to be {\bf absolutely continuous} if and only if it is differentiable almost everywhere, and $f(y)-f(x)=\int_x^yf'(z)dz$ for all $x,y\in [a,b]$. For another definition and the relationship between definitions, see Theorem 7.18 of Rudin (1987).} 4) the graph of $x$ is included in $U$, and 5) $\dot{x}(t)=g(t,x(t))$ for almost every $t\in I$. Let $x:I\to \mathbb{R}^N$ and $y:J\to \mathbb{R}^N$ be two solutions. Then, we say that $x$ is an {\bf extension} of $y$ if $J\subset I$ and $y(t)=x(t)$ for all $t\in J$. A solution $x:I\to \mathbb{R}^N$ is called a {\bf nonextendable solution} if there is no extension except $x$ itself. The next fact is well known, and thus we omit the proof.\footnote{See, for example, Theorems 1.1 and 3.1 in chapter 2 of Hartman (1997).}

\vspace{12pt}
\noindent
{\bf Fact 3}. Suppose that $g$ is locally Lipschitz. Then, for every interval $I$ including $t_0$, there exists at most one solution to (\ref{eq6.00}) defined on $I$, and if there exists a solution, it is continuously differentiable. In particular, there exists a unique nonextendable solution $x:I\to \mathbb{R}^N$ to (\ref{eq6.00}), where $I$ is open and $x(t)$ is continuously differentiable. Moreover, for every compact set $C\subset U$, there exist $t_1,t_2\in I$ such that if $t\in I$ and either $t<t_1$ or $t_2<t$, then $(t,x(t))\notin C$.

\vspace{12pt}
Next, consider the following parametrized ODE:
\begin{equation}\label{eq6.01}
\dot{x}(t)=h(t,x(t),y),\ x(t_0)=z,
\end{equation}
where $h:U\to \mathbb{R}^N$ and $U\subset \mathbb{R}\times \mathbb{R}^N\times \mathbb{R}^M$ is open. We assume that $h$ is locally Lipschitz. Fix $(y,z)$ such that $(t_0,z,y)\in U$. Then, (\ref{eq6.01}) can be seen as (\ref{eq6.00}), where $g(t,x)=h(t,x,y)$ and $x^*=z$. Hence, we can define a nonextendable solution $x^{y,z}:I\to \mathbb{R}^N$ according to Fact 3. We write $x(t;y,z)=x^{y,z}(t)$, and call this function $x:(t,y,z)\mapsto x(t;y,z)$ the {\bf solution function} of (\ref{eq6.01}). The following fact is necessary, but is not particularly well-known; thus, we prove it in this paper.

\vspace{12pt}
\noindent
{\bf Fact 4}. Under the assumption that $h$ is locally Lipschitz, the domain of the solution function is open, and the solution function is locally Lipschitz.

\vspace{12pt}
\noindent
{\bf Proof of Fact 4}. First, we introduce a lemma.

\vspace{12pt}
\noindent
{\bf Lemma 1} (Gronwall's inequality).\footnote{The first inequality of this lemma is famous and introduced by many textbooks. See, for example, problem 5.2.7 of Karatzas and Shreve (1998). However, we also need the second inequality in many situations, and there is no readable proof of this inequality in published textbooks. We provide the proof of these inequalities for readability and to make the present paper self-contained.} Suppose that $g:[t_0,t_1]\to \mathbb{R}$ is continuous, and
\[g(t)\le \int_{t_0}^tAg(s)ds+B(t),\]
for almost every $t\in [t_0,t_1]$, where $A>0$ and $B(t)$ is an integrable function on $[t_0,t_1]$. Then, for almost every $t\in [t_0,t_1]$,
\[g(t)\le B(t)+A\int_{t_0}^te^{A(t-s)}B(s)ds.\]
In particular, if $B(t)=C(t-t_0)$ for some constant $C$, then for every $t\in [t_0,t_1]$,
\[g(t)\le \frac{C}{A}(e^{A(t-t_0)}-1).\]

\vspace{12pt}
\noindent
{\bf Proof}. First, for almost every $t\in [t_0,t_1]$,
\[\frac{d}{dt}\left(e^{-At}\int_{t_0}^tg(s)ds\right)=e^{-At}\left(g(t)-\int_{t_0}^tAg(s)ds\right)\le e^{-At}B(t).\]
Integrating both sides, we obtain
\[e^{-At}\int_{t_0}^tg(s)ds\le \int_{t_0}^te^{-As}B(s)ds,\]
and thus,
\[g(t)\le A\int_{t_0}^tg(s)ds+B(t)\le B(t)+Ae^{At}\int_{t_0}^te^{-As}B(s)ds,\]
which implies that the first inequality holds almost everywhere. If $B(t)=C(t-t_0)$, then by continuity, the above inequality holds everywhere, and integration by parts yields
\begin{align*}
g(t)\le&~C(t-t_0)+AC\int_{t_0}^te^{A(t-s)}(s-t_0)ds\\
=&~C(t-t_0)+AC\left[-\frac{1}{A}e^{A(t-s)}(s-t_0)\right]_{t_0}^t+C\int_{t_0}^te^{A(t-s)}ds\\
=&~C\int_{t_0}^te^{A(t-s)}ds=\frac{C}{A}(e^{A(t-t_0)}-1),
\end{align*}
as desired. This completes the proof. $\blacksquare$

\vspace{12pt}
Let $V\subset \mathbb{R}\times\mathbb{R}^M\times \mathbb{R}^N$ be the domain of the solution function $x(t;y,z)$. Choose any $(t^*,y,z)\in V$. By Fact 3, there exists an open interval $I$ such that $t\mapsto x(t;y,z)$ is a nonextendable solution defined on $I$, and $t^*\in I$. Choose $t_1,t_2\in I$ such that $t_1<\min\{t^*,t_0\}\le \max\{t^*,t_0\}<t_2$. Consider the following differential equation:
\begin{equation}\label{eq6.02}
\dot{x}(t)=h(t,x(t)+z'-z,y'),\ x(t_0)=z.
\end{equation}
If $z'=z$ and $y'=y$, then $x^*:t\mapsto x(t;y,z)$ is a solution to (\ref{eq6.02}). Choose $a,b>0$ sufficiently small and define
\[\Pi(a,b)=\{(t,x,y',z')|t\in [t_1,t_2],\ \|x-x^*(t)\|\le a,\ \|y'-y\|\le b,\ \|z'-z\|\le b\},\]
\[\Pi'(a,b)=\{(t,x+z'-z,y')|(t,x,y',z')\in \Pi(a,b)\}.\]
By definition, $\Pi(a,b)$ and $\Pi'(a,b)$ are compact. We assume that $a,b$ are sufficiently small that $\Pi'(a,b)\subset U$. Because $h$ is locally Lipschitz, there exists $L>0$ such that, for every $(t_1',z_1',y_1'), (t_2',z_2',y_2')\in \Pi'(a,b)$,
\[\|h(t_1',z_1',y_1')-h(t_2',z_2',y_2')\|\le L[|t_1'-t_2'|+\|z_1'-z_2'\|+\|y_1'-y_2'\|].\]
Suppose that $\|y'-y\|\le b,\ \|z'-z\|\le b$, and define $t^+_2(d',c^*)$ as the supremum of the set of all $t\in ]t_0,t_2]$ such that there exists a solution $\tilde{x}:[t_0,t]\to \mathbb{R}^N$ to (\ref{eq6.02}) and $(s,\tilde{x}(s),y',z')\in \Pi(a,b)$ for all $s\in [t_0,t]$. By Fact 3, we have that $t^+_2(y',z')>t_0$ and there exists a solution $\tilde{x}:[t_0,t^+_2(y',z')]\to \mathbb{R}^N$ to (\ref{eq6.02}). If $t\in [t_0,t^+_2(y',z')]$, then
\begin{align*}
\|\tilde{x}(t)-x^*(t)\|\le&~\int_{t_0}^t\|h(s,\tilde{x}(s)+z'-z,y')-h(s,x^*(s),y)\|ds\\
\le&~\int_{t_0}^t\|h(s,\tilde{x}(s)+z'-z,y')-h(s,x^*(s)+z'-z,y')\|ds\\
 &~+\int_{t_0}^t\|h(s,x^*(s)+z'-z,y')-h(s,x^*(s),y)\|ds\\
\le&~\int_{t_0}^tL\|\tilde{x}(s)-x^*(s)\|ds+L(\|y'-y\|+\|z'-z\|)(t-t_0).
\end{align*}
Therefore, by Lemma 1,
\begin{equation}\label{eq6.03}
\|\tilde{x}(t)-x^*(t)\|\le (\|y'-y\|+\|z'-z\|)(e^{L(t-t_0)}-1).
\end{equation}
Choose $b'\in ]0,b[$ sufficiently small that
\begin{equation}\label{beta}
b'(e^{L(t_2-t_0)}-1)<a.
\end{equation}
Suppose that $\|y'-y\|+\|z'-z\|\le b'$ and $t_2^+(y',z')<t_2$. Because $\tilde{x}$ is defined at $t^+_2(y',z')$, we have that $(t^+_2(y',z'),\tilde{x}(t^+_2(y',z')),y',z')\in \Pi(a,b)$. By (\ref{eq6.03}) and (\ref{beta}), $\|\tilde{x}(t^+_2(y',z'))-x^*(t^+_2(y',z'))\|<a$, which contradicts the definition of $t^+_2(y',z')$. Therefore, if $\|y'-y\|+\|z'-z\|\le b'$, then $\tilde{x}(t)$ is defined on $[t_0,t_2]$, and if $t\in [t_0,t_2]$, then
\[\|\tilde{x}(t)-x^*(t)\|\le (\|y'-y\|+\|z'-z\|)(e^{L(t_2-t_1)}-1).\]
By a symmetric argument, we have that if $b'\in ]0,b[$ is so small that
\begin{equation}\label{beta2}
b'(e^{L(t_0-t_1)}-1)<a,
\end{equation}
and $\|y'-y\|+\|z'-z\|\le b'$, then (\ref{eq6.02}) has a solution $\tilde{x}$ defined on $[t_1,t_0]$, and if $t\in [t_1,t_0]$,
\[\|\tilde{x}(t)-x^*(t)\|\le (\|y'-y\|+\|z'-z\|)(e^{L(t_2-t_1)}-1).\]
Clearly,
\[x(t;y',z')=\tilde{x}(t)+z'-z,\]
and thus the domain $V$ of the solution function $x$ includes 
\[[t_1,t_2]\times \{(y',z')|\|y'-y\|+\|z'-z\|\le b'\},\]
which is a neighborhood of $(t^*,y,z)$. Moreover,
\[\max_{t\in [t_0,t_1]}\|x(t;y',z')-x(t;y,z)\|\le (\|y'-y\|+\|z'-z\|)e^{L(t_2-t_1)},\]
and thus Fact 1 implies that the solution function is locally Lipschitz. This completes the proof. $\blacksquare$

\vspace{12pt}
We now present the formula for the solution to linear differential equations. Consider the following ODE:
\[\dot{x}(t)=a(t)x(t),\]
where $I$ is an interval including $t_0$ and $a:I\to \mathbb{R}$ is a bounded measurable function on $I$. Then, the solution to the above equation is as follows:
\[x(t)=x(t_0)e^{\int_{t_0}^ta(s)ds}.\]
For a proof, see Theorem 1 of section 0.4 in Ioffe and Tikhomirov (1979).

Finally, we note a partial differential equation that appears in consumer theory. 

\vspace{12pt}
\noindent
{\bf Fact 5}. Let $f$ be a continuous CoD and $\succsim$ be a weak order such that $f=f^{\succsim}$. Define
\[E^x(p)=\inf\{p\cdot y|y\succsim x\}.\]
Then, the function $E^x$ is concave and continuous on $\mathbb{R}^n_{++}$. In addition, suppose that $f$ satisfies Walras' law and $x=f(p,m)$. Then, $E^x(p)=m$ and $E^x(q)>0$ for all $q\in \mathbb{R}^n_{++}$. Moreover, the function $E^x$ is continuously differentiable, and for every $q\in \mathbb{R}^n_{++}$,
\begin{equation}\label{Shephard}
\nabla E^x(q)=f(q,E^x(q)).
\end{equation}

\vspace{12pt}
This function $E^x$ is usually called the {\bf expenditure function}, and equality (\ref{Shephard}) is called {\bf Shephard's lemma}. For a proof, see Lemma 1 of Hosoya (2020).

\subsection{Proof of Proposition 1}
First, consider the following parametrized ODE:
\begin{equation}\label{eq6.1}
\dot{c}(t)=f((1-t)p+tq,c(t))\cdot (q-p),\ c(0)=w,
\end{equation}
and let $c(t;p,q,w)$ denote the solution function of (\ref{eq6.1}). We introduce two lemmas.

\vspace{12pt}
\noindent
{\bf Lemma 2}. Let $U=\mathbb{R}^n_{++}\times\mathbb{R}_{++}$. Choose any $(p,m)\in U$. Suppose that $W\subset U$ and the Lebesgue measure of $U\setminus W$ is zero. Moreover, suppose that $q\in \mathbb{R}^n_{++}$ and there exists $i^*\in \{1,...,n\}$ such that $q_{i^*}\neq p_{i^*}$, and that the domain of the solution function $c(t;p,q,w)$ of the ODE $(\ref{eq6.1})$ includes $[0,t^*]\times P_1^*\times P_2^*$ for $t^*>0$, where $P_1$ is a bounded open neighborhood of $q$, $P_2$ is a bounded open neighborhood of $m$, and $P_j^*$ denotes the closure of $P_j$. For every $(t,\tilde{r},w)\subset \mathbb{R}^{n+1}$ such that $t\in [0,t^*]$, $r\in P_1$ for
\[r_i=\begin{cases}
\tilde{r}_i & \mbox{if }i<i^*,\\
q_i & \mbox{if }i=i^*,\\
\tilde{r}_{i-1} & \mbox{if }i>i^*,
\end{cases}\]
 and $w\in P_2$, define
\[\xi(t,\tilde{r},w)=((1-t)p+tr,c(t;p,r,w)).\]
Then, the Lebesgue measure of $\xi^{-1}(U\setminus W)$ is also zero.

\vspace{12pt}
\noindent
{\bf Proof}. Without loss of generality, we assume that $i^*=n$. Throughout the proof of Lemma 2, we use the following notation. If $r\in \mathbb{R}^n$, then $\tilde{r}=(r_1,...,r_{n-1})\in \mathbb{R}^{n-1}$. Conversely, if $\tilde{r}\in \mathbb{R}^{n-1}$, then $r=(r_1,...,r_{n-1},q_n)$.

Let $\tilde{P}_1=\{\tilde{r}\in\mathbb{R}^{n-1}|r\in P_1\}$ and $\hat{P}_1$ be the closure of $\tilde{P}_1$. Although the actual domain of $\xi$ is $[0,t^*]\times \tilde{P}_1\times P_2$, throughout this proof, we consider that the domain of $\xi$ is $[0,t^*]\times \hat{P}_1\times P_2^*$. We show that $\xi$ is one-to-one on the set $]0,t^*]\times \hat{P}_1\times P_2^*$. Suppose that $t_1\neq 0\neq t_2$ and $\xi(t_1,\tilde{r}_1,w_1)=\xi(t_2,\tilde{r}_2,w_2)=(v,c)$. Because $v_n=(1-t_1)p_n+t_1q_n=(1-t_2)p_n+t_2q_n$ and $p_n\neq q_n$, we have that $t_1=t_2$. Because $v_i=(1-t_1)p_i+t_1r_{1i}=(1-t_1)p_i+t_1r_{2i}$ and $t_1\neq 0$, we have that $r_{1i}=r_{2i}$, and thus $\tilde{r}_1=\tilde{r}_2$. Therefore, it suffices to show that $c(t;p,r,w)$ is increasing in $w$. Suppose that $w_1<w_2$ and $c(t;p,r,w_1)\ge c(t;p,r,w_2)$. Because $c(0;p,r,w_1)=w_1<w_2=c(0;p,r,w_2)$, by the intermediate value theorem, there exists $s\in [0,t]$ such that $c(s;p,r,w_1)=c(s;p,r,w_2)$. Then, by Fact 3, we have $w_1=c(0;p,r,w_1)=c(0;p,r,w_2)=w_2$, which is a contradiction.

Next, define
\[V^{\ell}=\xi([\ell^{-1}t^*,t^*]\times \hat{P}_1\times P_2^*).\]
We show that $\xi^{-1}$ is Lipschitz on $V^{\ell}$. Define
\[t(v)=\frac{v_n-p_n}{q_n-p_n},\]
\[\tilde{r}(v)=\frac{1}{t(v)}[(t(v)-1)\tilde{p}+\tilde{v}].\]
Suppose that $(v_1,c_1), (v_2,c_2)\in V^{\ell}$ and $(v_j,c_j)=\xi(t_j,\tilde{r}_j,w_j)$. Then, we have $t_j=t(v_j)$ and $\tilde{r}_j=\tilde{r}(v_j)$. Clearly, the functions $t(v)$ and $\tilde{r}(v)$ are Lipschitz on $V^{\ell}$. Next, consider the following ODE:
\[\dot{d}(s)=f((1-(s+t-t_2))p+(s+t-t_2)r(v),d(s))\cdot (r(v)-p),\ d(t_2)=c.\]
Let $d(s;t,v,c)$ be the solution function of this ODE. If $(v,c)=\xi(t,\tilde{r},w)$ for some $(t,\tilde{r},w)\in [{\ell}^{-1}t^*,t^*]\times \hat{P}_1\times P_2^*$, then $d(s;t,v,c)=c(s+t-t_2;p,r,w)$. Moreover, the set
\[\{(t,v,c)|t\in [\ell^{-1}t^*,t^*],\ (v,c)=\xi(t,\tilde{r},w)\mbox{ for some }(\tilde{r},w)\in \hat{P}_1\times P_2^*\}\]
is compact, and by Fact 4, $(t,v,c)\mapsto d(t_2-t;t,v,c)$ is Lipschitz on this set. Therefore,
\begin{align*}
|w_1-w_2|=&~|d(t_2-t_1;t_1,v_1,c_1)-d(t_2-t_2;t_2,v_2,c_2)|\\
\le&~L[|t_1-t_2|+\|(v_1,c_1)-(v_2,c_2)\|]\\
=&~L[|t(v_1)-t(v_2)|+\|(v_1,c_1)-(v_2,c_2)\|]\\
\le&~L(M+1)\|(v_1,c_1)-(v_2,c_2)\|,
\end{align*}
where $L,M>0$ are some constants, and therefore our claim is correct.

Now, recall that the Lebesgue measure of $U\setminus W$ is zero. Because $\xi^{-1}$ is Lipschitz on $V^{\ell}$, we have that the Lebesgue measure of
\[\xi^{-1}(V^{\ell}\cap (U\setminus W))\]
is zero. Therefore, the Lebesgue measure of
\[\cup_{\ell}\xi^{-1}(V^{\ell}\cap (U\setminus W))\]
is also zero. Clearly, the Lebesgue measure of
\[\xi^{-1}(U\setminus W)\setminus \left(\cup_{\ell}\xi^{-1}(V^{\ell}\cap (U\setminus W))\right)\]
is zero, because this set is included in $\{0\}\times \hat{P}_1\times P_2^*$. This completes the proof of Lemma 2. $\blacksquare$

\vspace{12pt}
\noindent
{\bf Lemma 3}. Choose any $(p,m)\in \mathbb{R}^n_{++}$. Then, there exists a solution $E:\mathbb{R}^n_{++}\to \mathbb{R}_{++}$ to the partial differential equation
\begin{equation}\label{eq6.1.1}
\nabla E(q)=f(q,E(q)),\ E(p)=m,
\end{equation}
if and only if the domain of the solution function of (\ref{eq6.1}) includes $[0,1]\times \{p\}\times \mathbb{R}^n_{++}\times \{m\}$. Moreover, in this case, for each $q\in \mathbb{R}^n_{++}$,
\[E(q)=c(1;p,q,m).\]

\vspace{12pt}
\noindent
{\bf Proof}. Suppose that a solution $E:\mathbb{R}^n_{++}\to \mathbb{R}_{++}$ to (\ref{eq6.1.1}) exists. Choose any $q\in \mathbb{R}^n_{++}$. Let $d(t)=E((1-t)p+tq)$. Then, $d(0)=E(p)=m$ and
\[\dot{d}(t)=f((1-t)p+tq,d(t))\cdot (q-p),\]
and by the uniqueness of the solution to an ODE (Fact 3), we have that $d(t)\equiv c(t;p,q,m)$. Hence, the domain of the solution function $c$ includes $[0,1]\times \{p\}\times \mathbb{R}^n_{++}\times \{m\}$, and moreover, $E(q)=d(1)=c(1;p,q,m)$.

We show that the converse is also true. Suppose that the domain of the solution function $c$ includes $[0,1]\times \{p\}\times \mathbb{R}^n_{++}\times \{m\}$. Define $E(q)=c(1;p,q,m)$. We show that $E(q)$ is a solution to (\ref{eq6.1.1}).

First, let $\Delta^*$ be the set of all $(q,w)$ such that $f$ is differentiable and $S_f$ is symmetric and negative semi-definite at $((1-t)p+tq,c(t;p,q,w))$ for almost every $t\in [0,1]$, and the mapping $\tilde{r}\mapsto c(t;p,\tilde{r},q_n,w)$ is differentiable at $\tilde{r}=(q_1,...,q_{n-1})$ for almost every $t\in [0,1]$. Suppose that $(q,w)\in \Delta^*$ and let $e_i$ denote the $i$-th unit vector. Then, for each $i\in \{1,...,n-1\}$,\footnote{In this proof, we frequently abbreviate several variables for simplicity.}
\begin{align*}
&~\lim_{h\to 0}\frac{c(t;p,q+he_i,w)-c(t;p,q,w)}{h}\\
=&~\lim_{h\to 0}\frac{1}{h}\times\\
&~\left[\int_0^tf((1-s)p+s(q+he_i),c(s;p,q+he_i,w))\cdot (q+he_i-p)ds\right.\\
&~-\left.\int_0^tf((1-s)p+sq,c(s;p,q,w))\cdot (q-p)ds\right]\\
=&~\int_0^t\left[f_i+\sum_{j=1}^n\left[s\frac{\partial f_j}{\partial p_i}+\frac{\partial f_j}{\partial m}\frac{\partial c}{\partial q_i}\right](q_j-p_j)\right]ds,
\end{align*}
by the dominated convergence theorem, and thus $\frac{\partial c}{\partial q_i}(t;p,q,w)$ is defined for all $t\in [0,1]$ and is absolutely continuous in $t$. Define the following absolutely continuous function
\[\varphi(t)=\frac{\partial c}{\partial q_i}(t;p,q,w)-tf_i((1-t)p+tq,c(t;p,q,w)).\]
By the above evaluation and the symmetry of the Slutsky matrix, we have that for almost all $t\in [0,1]$,
\begin{align*}
\dot{\varphi}(t)=&~f_i+\sum_{j=1}^n\left[t\frac{\partial f_j}{\partial p_i}+\frac{\partial f_j}{\partial m}\frac{\partial c}{\partial q_i}\right](q_j-p_j)-f_i-t\sum_{j=1}^n\left[\frac{\partial f_i}{\partial p_j}+\frac{\partial f_i}{\partial m}f_j\right](q_j-p_j)\\
=&~t\sum_{j=1}^n\left[\frac{\partial f_j}{\partial p_i}-\frac{\partial f_i}{\partial p_j}-\frac{\partial f_i}{\partial m}f_j\right](q_j-p_j)+\frac{\partial c}{\partial q_i}\sum_{j=1}^n\frac{\partial f_j}{\partial m}(q_j-p_j)\\
=&~\varphi(t)\sum_{j=1}^n\frac{\partial f_j}{\partial m}(q_j-p_j)\\
\equiv&~a(t)\varphi(t),
\end{align*}
where $a(t)$ is some bounded measurable function. By the formula for the solution to linear ODEs, we have that
\[\varphi(t)=\varphi(0)e^{\int_0^ta(s)ds}.\]
However, we can easily check that $\varphi(0)=0$, and thus $\varphi(t)\equiv 0$. In particular, $\varphi(1)=0$, and thus
\[\frac{\partial c}{\partial q_i}(1;p,q,w)=f_i(q,c(1;p,q,w)).\]

Second, suppose that $q_n\neq p_n$ and $i\in \{1,...,n-1\}$. By Lemma 2 and Fubini's theorem, there exist $\delta>0$ and a sequence $(q^k,w^k)$ on $\Delta^*$ such that $q^k\to q, w^k\to m$ as $k\to \infty$, and for every $k$, $i\in \{1,...,n-1\}$ and almost every $h\in ]-\delta,\delta[$, $(q^k+he_i,w^k)\in \Delta^*$. Then, for every $h\in ]-\delta,\delta[$,
\[c(1;p,q^k+he_i,w^k)-c(1;p,q^k,w^k)=\int_0^hf_i(q^k+se_i,c(1;p,q^k+se_i,w^k))ds.\]
By the dominated convergence theorem, we have that
\[E(q+he_i)-E(q)=\int_0^hf_i(q+se_i,E(q+se_i))ds,\]
which implies that
\[\frac{\partial E}{\partial q_i}(q)=f_i(q,E(q)).\]

Third, suppose that $q_n=p_n$ and $i\in \{1,...,n-1\}$. Let $e=(1,1,...,1)$ and define $q^k=q+k^{-1}e$. Then, $q_n^k\neq p_n$, and thus, for every $h\in ]-q_i,q_i[$,
\[E(q^k+he_i)-E(q^k)=\int_0^hf_i(q^k+se_i,E(q^k+se_i))ds,\]
and by the dominated convergence theorem,
\[E(q+he_i)-E(q)=\int_0^hf_i(q+se_i,E(q+se_i))ds,\]
which implies that
\[\frac{\partial E}{\partial q_i}(q)=f_i(q,E(q)).\]
In summary, we obtain the following: for every $q\in \mathbb{R}^n_{++}$ and $i\in \{1,...,n-1\}$,
\begin{equation}\label{eq6.3}
\frac{\partial E}{\partial q_i}(q)=f_i(q,E(q)).
\end{equation}
Replacing the role of $n$ with that of $1$ and repeating the above arguments, we can show that (\ref{eq6.3}) holds for $i=n$, and thus $\nabla E(q)=f(q,E(q))$. This completes the proof. $\blacksquare$

\vspace{12pt}
We now complete the preparation for proving Proposition 1. We separate the proof of Proposition 1 into ten steps.

\vspace{12pt}
\noindent
{\bf Step 1}. Suppose that $t^*>0$ and the domain of the solution function $c(t;p,q,w)$ of (\ref{eq6.1}) includes $[0,t^*]\times \{(p,q,m)\}$. Define $p(t)=(1-t)p+tq$ and $x(t)=f(p(t),c(t;p,q,m))$. Then, $p\cdot x(t^*)\ge m$ and $p(t^*)\cdot x(0)\ge c(t^*;p,q,m)$.

\vspace{12pt}
\noindent
{\bf Proof of Step 1}. We prove only the former claim, because the latter claim can be shown symmetrically. Define $p(t,r)=(1-t)p+tr$, and let $\Delta(t^*)$ be the set of all $(r,w)$ such that the domain of $t\mapsto c(t;p,r,w)$ includes $[0,t^*]$, and for almost every $t\in [0,t^*]$, $f$ is differentiable and $S_f$ is symmetric and negative semi-definite at $(p(t,r),c(t;p,r,w))$. By Lemma 2 and Fubini's theorem, there exists a sequence $(q^k,w^k)$ on $\Delta(t^*)$ that converges to $(q,m)$ as $k\to \infty$. Define $d(t)=p\cdot x(t)$ and $d^k(t)=p\cdot f(p(t,q^k),c(t;p,q^k,w^k))$. Then, $d^k$ is absolutely continuous, and for almost all $t\in [0,t^*]$, 
\[\dot{d}^k(t)=p^TS_f(p(t,q^k),c(t;p,q^k,w^k))(q^k-p).\]
Now, differentiating both sides of Walras' law, we obtain
\[(p(t,q^k))^TS_f(p(t,q^k),c(t;p,q^k,w^k))(q^k-p)=0.\]
Subtracting the latter from the former, we have that
\[\dot{d}^k(t)=-t(q^k-p)^TS_f(p(t,q^k),c(t;p,q^k,w^k))(q^k-p)\ge 0,\]
and thus, $d^k(t)$ is nondecreasing. Because $d^k(t)\to d(t)$ for every $t$, we have that $d(t)$ is also nondecreasing, and thus
\[p\cdot x(t^*)=d(t^*)\ge d(0)=m,\]
which completes the proof of Step 1. $\blacksquare$

\vspace{12pt}
\noindent
{\bf Step 2}. The domain of the solution function $c(t;p,q,m)$ includes $[0,1]\times \mathbb{R}^n_{++}\times \mathbb{R}^n_{++}\times \mathbb{R}_{++}$.

\vspace{12pt}
\noindent
{\bf Proof of Step 2}. Suppose not. Then, there exist $p,q\in \mathbb{R}^n_{++}$ and $m\in\mathbb{R}_{++}$ such that $c(t;p,q,m)$ is defined only on $[0,t^*[$, where $t^*\le 1$. Let $p(t)=(1-t)p+tq$ and $x=f(p,m)$. By Fact 3, we have that the trajectory of the function $(p(t),c(t;p,q,m))$ is excluded from any compact set in $\mathbb{R}^n_{++}\times\mathbb{R}_{++}$ as $t\to t^*$, and thus either $\limsup_{t\to t^*}c(t;p,q,m)=+\infty$ or $\liminf_{t\to t^*}c(t;p,q,m)=0$. By Step 1, $\max\{p\cdot x,q\cdot x\}\ge p(t)\cdot x\ge c(t;p,q,m)$ for every $t\in [0,t^*[$, and therefore, there exists an increasing sequence $(t^k)$ such that $t^k\to t^*$ and $c(t^k;p,q,m)\to 0$ as $k\to \infty$. Define $x^k=f(p(t^k),c(t^k;p,q,m))$. Because $p\cdot x^k\ge m=p\cdot x$ and $p(t^k)\cdot x\ge c(t^k;p,q,m)=p(t^k)\cdot x^k$, we have that $q\cdot x\ge q\cdot x^k$. Hence, $(x^k)$ is a sequence in the following compact set
\[\{y\in \Omega|q\cdot y\le q\cdot x\}.\]
Therefore, without loss of generality, we can assume that $x^k\to x^*$ as $k\to \infty$. Because $p\cdot x^*\ge m$, we have that $x^*\gneq 0$. However,
\[0<p(t^*)\cdot x^*=\lim_{k\to \infty}p(t^k)\cdot x^k=\lim_{k\to \infty}c(t^k;p,q,m)=0,\]
which is a contradiction. This completes the proof of Step 2. $\blacksquare$

\vspace{12pt}
\noindent
{\bf Step 3}. For all $t\in [0,1]$, $c(1-t;p,q,m)=c(t;q,p,c(1;p,q,m))$. Moreover, if $m>m'$, then $c(t;p,q,m)>c(t;p,q,m')$ for every $t\in [0,1]$.

\vspace{12pt}
\noindent
{\bf Proof of Step 3}. First,
\[\frac{d}{dt}c(1-t;p,q,m)=f((1-t)q+tp,c(1-t;p,q,m))\cdot (p-q),\]
and thus, by the uniqueness of the solution to an ODE (Fact 3), we have that
\[c(1-t;p,q,m)=c(t;q,p,c(1;p,q,m)).\]
Next, suppose that $c(t;p,q,m)\le c(t;p,q,m')$ for some $t\in [0,1]$. By the intermediate value theorem, there exists $s\in [0,1]$ such that $c(s;p,q,m)=c(s;p,q,m')$. Again, by the uniqueness of the solution to an ODE, we have that
\[m=c(0;p,q,m)=c(0;p,q,m')=m',\]
which is a contradiction. This completes the proof of Step 3. $\blacksquare$

\vspace{12pt}
\noindent
{\bf Step 4}. Suppose that $x\neq y,\ x=f(p,m),\ y=f(q,w)$, and $w\ge c(1;p,q,m)$. Then, $p\cdot y>m$.

\vspace{12pt}
\noindent
{\bf Proof of Step 4}. First, define $m^*=c(1;q,p,w)$.\footnote{Note that, this is equivalent to $w=c(1;p,q,m^*)$.} By Step 3, we have that $c(t;q,p,w)=c(1-t;p,q,m^*)$, and thus $m^*\ge m$. Moreover, $w>c(1;p,q,m)$ if and only if $m^*>m$.

Define $p(t)=(1-t)p+tq$ and $d(t)=p\cdot f(p(t),c(t;p,q,m^*))$. We have already shown in the proof of Step 1 that $d(t)$ is nondecreasing. Therefore, if $m^*>m$, then
\[p\cdot y=p\cdot f(q,w)=d(1)\ge d(0)=m^*>m,\]
as desired. Thus, we hereafter assume that $m^*=m$. In this regard, we have that $w=c(1;p,q,m)$, and $c(1-t;q,p,w)=c(t;p,q,m)$.

Define $\Delta^*$ as the set of all $(r,c)$ such that $f$ is differentiable and $S_f$ is symmetric and negative semi-definite at $((1-t)p+tr,c(t;p,r,c))$ for almost all $t\in [0,1]$. By Lemma 2 and Fubini's theorem, there exists a sequence $(q^k,w^k)$ on $\Delta^*$ such that $(q^k,w^k)\to (q,m)$ as $k\to \infty$. Let $2\varepsilon=\|x-y\|$, and define $p^k(t)=(1-t)p+tq^k$ and $x^k(t)=f(p^k(t),c(t;p,q^k,w^k))$. Then, $x^k(1)\to y$ and $x^k(0)\to x$ as $k\to \infty$, and thus, without loss of generality, we can assume that $\|x^k(1)-x^k(0)\|\ge \varepsilon$ for all $k$. By assumption, $x^k(t)$ is a Lipschitz function defined on $[0,1]$. If $f$ is differentiable at $(p^k(t),c(t;p,q^k,w^k))$, then define $S_t^k=S_f(p^k(t),c(t;p,q^k,w^k))$. By assumption, $S_t^k$ can be defined and is symmetric and negative semi-definite for almost all $t\in [0,1]$. Because $S_t^k$ is symmetric and negative semi-definite, there exists a positive semi-definite matrix $A_t^k$ such that $S_t^k=-(A_t^k)^2$. Moreover, the operator norm $\|A_t^k\|$ is equal to $\sqrt{\|S_t^k\|}$.\footnote{Because $S_t^k$ is symmetric, there exists an orthogonal transform $P$ such that
\[S_t^k=P^T\begin{pmatrix}
\lambda_1 & 0 & ... & 0\\
0 & \lambda_2 & ... & 0\\
\vdots & \vdots & \ddots & \vdots\\
0 & 0 & ... & \lambda_n
\end{pmatrix}P,\]
where $\lambda_i$ is some eigenvalue of $S_t^k$. Because $S_t^k$ is negative semi-definite, $\lambda_i\le 0$ for every $i$. Hence, if we define
\[A_t^k=P^T\begin{pmatrix}
\sqrt{-\lambda_1} & 0 & ... & 0\\
0 & \sqrt{-\lambda_2} & ... & 0\\
\vdots & \vdots & \ddots & \vdots\\
0 & 0 & ... & \sqrt{-\lambda_n}
\end{pmatrix}P,\]
then $S_t^k=-(A_t^k)^2$. Moreover, because the operator norm $\|S_t^k\|$ (resp. $\|A_t^k\|$) coincides with $\max_i|\lambda_i|$, (resp. $\max_i\sqrt{|\lambda_i|}$,) all our claims are correct.} Because $f$ is locally Lipschitz, there exists $L>0$ such that $\|S_t^k\|\le L$ for all $k$ and almost all $t\in [0,1]$. Define $d^k(t)=p\cdot x^k(t)$, and choose $\delta>0$ such that $\varepsilon^2>2L^2\delta\|q^k-p\|^2$ for every sufficiently large $k$. By the same arguments as in the proof of Step 1, we can show that
\[\dot{d}^k(t)=-t(q^k-p)^TS_t^k(q^k-p)=t\|A_t^k(q^k-p)\|^2,\]
\[\dot{x}^k(t)=S_t^k(q^k-p)\]
for almost all $t\in [0,1]$. Therefore,
\begin{align*}
\varepsilon^2\le \|x^k(1)-x^k(0)\|^2=&~\left\|\int_0^1\dot{x}^k(t)dt\right\|^2\le \int_0^1\|\dot{x}^k(t)\|^2dt=\int_0^1\|S_t^k(q^k-p)\|^2dt\\
\le&~\int_0^1\|A_t^k\|^2\|A_t^k(q^k-p)\|^2dt\le L\int_0^1\|A_t^k(q^k-p)\|^2dt\\
=&~L\int_0^{\delta}\|A_t^k(q^k-p)\|^2dt+L\int_{\delta}^1\frac{1}{t}\dot{d}^k(t)dt\\
\le&~L^2\delta\|q^k-p\|^2+\frac{L}{\delta}(d^k(1)-d^k(\delta)),
\end{align*}
and thus,
\[\frac{\delta \varepsilon^2}{2L}\le d^k(1)-d^k(\delta).\]
Letting $k\to \infty$, we have that
\[p\cdot y=d(1)\ge d(\delta)+\frac{\delta \varepsilon^2}{2L}>d(0)=m,\]
as desired. This completes the proof of Step 4. $\blacksquare$

\vspace{12pt}
\noindent
{\bf Step 5}. If $x\neq y,\ x=f(p,m),\ y=f(q,w)$, and $p\cdot y\le m$, then $q\cdot x>w$.\footnote{In other words, $f$ satisfies the {\bf weak axiom of revealed preference}.}

\vspace{12pt}
\noindent
{\bf Proof of Step 5}. By the contrapositive of Step 4, we have that $c(1;p,q,m)>w=c(0;q,p,w)$. By Step 3, $m=c(0;p,q,m)>c(1;q,p,w)$. By Step 4, we obtain that $q\cdot x>w$, as desired. This completes the proof of Step 5. $\blacksquare$

\vspace{12pt}
\noindent
{\bf Step 6}. For every $q\in \mathbb{R}^n_{++}$, $c(1;q,\bar{p},c(1;p,q,m))=c(1;p,\bar{p},m)$.

\vspace{12pt}
\noindent
{\bf Proof of Step 6}. Define $p(t)=(1-t)p+tq$, $m^*=c(1;p,\bar{p},m)$ and $w(t)=c(1;\bar{p},p(t),m^*)$. By Step 3, $c(1-t;p,\bar{p},m)=c(t;\bar{p},p,m^*)$, and thus $w(0)=c(0;p,\bar{p},m)=m$. Moreover, by Lemma 3 and Step 2, $E(r)=c(1;\bar{p},r,m^*)$ satisfies the following differential equation:
\[\nabla E(r)=f(r,E(r)).\]
Therefore,
\begin{align*}
\dot{w}(t)=&~\frac{d}{dt}c(1;\bar{p},(1-t)p+tq,m^*)=\frac{d}{dt}E((1-t)p+tq)\\
=&~f((1-t)p+tq,E((1-t)p+tq))\cdot (q-p)\\
=&~f((1-t)p+tq,w(t))\cdot (q-p).
\end{align*}
By the uniqueness of the solution to an ODE, we have that
\[w(t)=c(t;p,q,m)\]
for all $t\in [0,1]$. Now, define $m^+=c(1;p,q,m)$. Then,
\[c(1;\bar{p},q,m^*)=w(1)=m^+.\]
By Step 3, we have $c(1-t;\bar{p},q,m^*)=c(t;q,\bar{p},m^+)$, and thus
\[m^*=c(1;q,\bar{p},m^+),\]
as desired. This completes the proof of Step 6. $\blacksquare$

\vspace{12pt}
\noindent
{\bf Step 7}. Suppose that $x=f(p,m)=f(q,w)$. Then, $c(1;p,\bar{p},m)=c(1;q,\bar{p},w)$.

\vspace{12pt}
\noindent
{\bf Proof of Step 7}. Let $p(t)=(1-t)p+tq$ and $m(t)=(1-t)m+tw$. Suppose that $f(p(t),m(t))=y\neq x$ for some $t\in [0,1]$. By Walras' law, $p(t)\cdot y=m(t)=p(t)\cdot x$, and thus $p\cdot y>m$ and $q\cdot y>w$ by Step 5. However, this implies that $p(t)\cdot y>m(t)$, which is a contradiction. Therefore, $f(p(t),m(t))\equiv x$, and thus,
\[\dot{m}(t)=w-m=x\cdot (q-p)=f(p(t),m(t))\cdot (q-p).\]
By the uniqueness of the solution to an ODE,
\[m(t)=c(t;p,q,m),\]
and thus $w=c(1;p,q,m)$. Therefore, by Step 6, we have that $c(1;p,\bar{p},m)=c(1;q,\bar{p},w)$. This completes the proof of Step 7. $\blacksquare$

\vspace{12pt}
By Steps 2 and 7, we can define $u_{f,\bar{p}}(x)$ for all $x\in \Omega$, and our definition of $u_{f,\bar{p}}(x)$ is independent of the choice of $(p,m)\in f^{-1}(x)$.

\vspace{12pt}
\noindent
{\bf Step 8}. $f=f^{u_{f,\bar{p}}}$.

\vspace{12pt}
\noindent
{\bf Proof of Step 8}. Suppose that $x=f(p,m), y\neq x$, and $p\cdot y\le m$. If $y\notin R(f)$, then $u_{f,\bar{p}}(y)=0<u_{f,\bar{p}}(x)$. If $y=f(q,w)$ for some $(q,w)$, then the contrapositive of Step 4 implies that $c(1;p,q,m)>w$. By Step 3,
\[c(t;q,\bar{p},c(1;p,q,m))>c(t;q,\bar{p},w)\]
for every $t\in [0,1]$. By Step 6,
\[u_{f,\bar{p}}(x)=c(1;p,\bar{p},m)=c(1;q,\bar{p},c(1;p,q,m))>c(1;q,\bar{p},w)=u_{f,\bar{p}}(y).\]
Therefore, $x=f^{u_{f,\bar{p}}}(p,m)$, as desired. This completes the proof of Step 8. $\blacksquare$

\vspace{12pt}
\noindent
{\bf Step 9}. $u_{f,\bar{p}}$ is upper semi-continuous on $R(f)$.

\vspace{12pt}
\noindent
{\bf Proof of Step 9}. Suppose that $x=f(p,m)$ and $u_{f,\bar{p}}(x)<a$. By Fact 4, the solution function $c$ is continuous, and thus there exists $\varepsilon>0$ such that $c(1;p,\bar{p},m+\varepsilon)<a$. Define $y=f(p,m+\varepsilon)$. Then, the set
\[U=\{z\in R(f)|p\cdot z<m+\varepsilon\}\]
is a neighborhood of $x$ in the relative topology of $R(f)$, and for every $z\in U$, $u_{f,\bar{p}}(z)<u_{f,\bar{p}}(y)<a$. This completes the proof of Step 9. $\blacksquare$

\vspace{12pt}
\noindent
{\bf Step 10}. Suppose that $f=f^{\succsim}$ for some weak order $\succsim$, and $\succsim$ is upper semi-continuous on $R(f)$. Then, for every $x,y\in R(f)$,
\[x\succsim y\Leftrightarrow u_{f,\bar{p}}(x)\ge u_{f,\bar{p}}(y).\]

\vspace{12pt}
\noindent
{\bf Proof of Step 10}. First, choose any $x\in R(f)$ and suppose that $x=f(p,m)$. Define $z=f(\bar{p},u_{f,\bar{p}}(x))$, and let
\[E^x(q)=\inf\{q\cdot w|w\succsim x\}.\]
By Fact 5, we have that
\[\nabla E^x(q)=f(q,E^x(q)),\ E^x(p)=m.\]
Define $d(t)=E^x((1-t)p+t\bar{p})$. Then,
\[\dot{d}(t)=f((1-t)p+t\bar{p},d(t))\cdot (\bar{p}-p),\]
which implies that $d(t)=c(t;p,\bar{p},m)$ for every $t\in [0,1]$. In particular,
\[E^x(\bar{p})=d(1)=c(1;p,\bar{p},m)=u_{f,\bar{p}}(x).\]
Now, choose any $\varepsilon>0$. Then, there exists $w\in \Omega$ such that $p\cdot w<E^x(\bar{p})+\varepsilon$ and $w\succsim x$. Define $z^{\varepsilon}=f(\bar{p},E^x(\bar{p})+\varepsilon)$. Then, $z^{\varepsilon}\succsim w$, and thus $z^{\varepsilon}\succsim x$. Letting $\varepsilon\to 0$, by the upper semi-continuity of $\succsim$, we obtain that $z\succsim x$.

Next, define
\[E^z(q)=\inf\{q\cdot w|w\succsim z\}.\]
By the same arguments as above, we can show that $E^z(p)=m$, and thus $x\succsim z$. Hence, $x\sim f(\bar{p},u_{f,\bar{p}}(x))$ for all $x\in R(f)$.

Now, choose any $x,y\in R(f)$. Then,
\[x\succsim y\Leftrightarrow f(\bar{p},u_{f,\bar{p}}(x))\succsim f(\bar{p},u_{f,\bar{p}}(y))\Leftrightarrow u_{f,\bar{p}}(x)\ge u_{f,\bar{p}}(y),\]
as desired. This completes the proof of Step 10. $\blacksquare$

\vspace{12pt}
Steps 8--10 indicate that all of our claims in Proposition 1 are correct. This completes the proof. $\blacksquare$

\subsection{Proof of Corollary 1}
It is obvious that condition (ii) implies condition (i).

Suppose that condition (i) holds, and choose any $(p,m)\in\mathbb{R}^n_{++}\times\mathbb{R}_{++}$. Let $x=f(p,m)$ and define
\[E^x(q)=\inf\{q\cdot y|y\succsim x\}.\]
By Fact 5, this $E^x$ is a concave solution to (\ref{eq3.2}). Suppose that $E$ is another solution to (\ref{eq3.2}). For each $q\in \mathbb{R}^n_{++}$, define $c_1(t)=E^x((1-t)p+tq)$ and $c_2(t)=E((1-t)p+tq)$. Then,
\[\dot{c}_i(t)=f((1-t)p+tq,c_i(t))\cdot (q-p),\ c_i(0)=m,\]
and thus, by the uniqueness of the solution to an ODE (Fact 3), we have that $c_1\equiv c_2$. In particular, $E^x(q)=c_1(1)=c_2(1)=E(q)$, and thus $E=E^x$. Therefore, the solution is unique, and condition (iv) holds.

Suppose that condition (iv) holds. Choose any $(p,m)\in \mathbb{R}^n_{++}\times \mathbb{R}_{++}$ such that $f$ is differentiable at $(p,m)$. Let $E$ be a concave solution to (\ref{eq3.2}). By an easy calculation, we obtain
\[H_E(p)=S_f(p,m),\]
where $H_E(p)$ denotes the Hessian matrix of $E$ at $p$. Because $E$ is concave, $H_E(p)$ is negative semi-definite. Moreover, by extended Young's theorem,\footnote{See Lemma 3.2 of Hosoya (2021).} $H_E(p)$ is symmetric. Therefore, $f$ satisfies (S) and (NSD), and condition (iii) holds.

Finally, our Proposition 1 says that condition (iii) implies condition (ii). This completes the proof. $\blacksquare$

\subsection{Proof of Corollary 2}
Define
\[w_{f,\bar{p}}(x)=\begin{cases}
u_{f,\bar{p}}(x) & \mbox{if }x\in R(f),\\
\inf_{\varepsilon>0}\sup\{u_{f,\bar{p}}(y)|y\in R(f),\ \|y-x\|<\varepsilon\} & \mbox{if }x\notin R(f).
\end{cases}\]
Theorems 1 and 2 of Hosoya (2020) state the following facts: 1) $f=f^{w_{f,\bar{p}}}$, 2) $w_{f,\bar{p}}$ is upper semi-continuous, 3) $w_{f,\bar{p}}$ is continuous on $\mathbb{R}^n_{++}$ if and only if $f$ satisfies the C axiom, and 4) if $\succsim$ is an upper semi-continuous weak order on $\Omega$ such that $f=f^{\succsim}$, then for each $x,y\in \Omega$,
\[x\succsim y\Leftrightarrow w_{f,\bar{p}}(x)\ge w_{f,\bar{p}}(y).\]
We first show that $w_{f,\bar{p}}(x)=v_{f,\bar{p}}(x)$ for all $x\in \Omega$.

If $x\in \mathbb{R}^n_{++}$, then $w_{f,\bar{p}}(x)=u_{f,\bar{p}}(x)=v_{f,\bar{p}}(x)$.

Suppose that $x\notin R(f)$. Choose any $\varepsilon>0$, and suppose that $y\in R(f)$ and $\|y-x\|<\varepsilon$. Then, there exists $z\in \mathbb{R}^n_{++}$ such that $z\gg y$ and $\|z-x\|<\varepsilon$. If $z=f(p,m)$, then $p\cdot y<m$, and thus $u_{f,\bar{p}}(z)>u_{f,\bar{p}}(y)$. This indicates that
\[\sup\{u_{f,\bar{p}}(y)|y\in R(f),\ \|y-x\|<\varepsilon\}=\sup\{u_{f,\bar{p}}(y)|y\in\mathbb{R}^n_{++},\ \|y-x\|<\varepsilon\},\]
and thus, $v_{f,\bar{p}}(x)=w_{f,\bar{p}}(x)$.

Suppose that $x\in R(f)\setminus \mathbb{R}^n_{++}$. Let $e=(1,1,...,1)$ and define $x^k=x+k^{-1}e$. Then, $x^k\in \mathbb{R}^n_{++}$. It is easy to show that
\[\lim_{k\to \infty}u_{f,\bar{p}}(x^k)=v_{f,\bar{p}}(x).\]
Because $u_{f,\bar{p}}$ is upper semi-continuous on $R(f)$, we have that
\[\lim_{k\to \infty}u_{f,\bar{p}}(x^k)\le u_{f,\bar{p}}(x).\]
On the other hand, if $x^k=f(p^k,m^k)$, then $p^k\cdot x<m^k$, and thus, $u_{f,\bar{p}}(x^k)>u_{f,\bar{p}}(x)$. Therefore, we have that
\[\lim_{k\to \infty}u_{f,\bar{p}}(x^k)\ge u_{f,\bar{p}}(x).\]
Combining these inequalities, we have that
\[v_{f,\bar{p}}(x)=\lim_{k\to \infty}u_{f,\bar{p}}(x^k)=w_{f,\bar{p}}(x),\]
as desired. Hence, $v_{f,\bar{p}}(x)=w_{f,\bar{p}}(x)$.

The rest of the claim of this corollary is statement 3). If $v_{f,\bar{p}}$ is continuous, then $f=f^{\succsim}$ for a continuous weak order $\succsim$ defined as
\[x\succsim y\Leftrightarrow v_{f,\bar{p}}(x)\ge v_{f,\bar{p}}(y).\]
Let us show the converse. Suppose that there exists a continuous weak order $\succsim$ on $\Omega$ such that $f=f^{\succsim}$. Debreu (1954) showed that there exists a continuous function $u:\Omega\to \mathbb{R}$ that represents $\succsim$. By the above arguments, $v_{f,\bar{p}}$ also represents $\succsim$, and thus $v_{f,\bar{p}}$ and $u$ have the same order. On the other hand, in the proof of Theorem 1 of Hosoya (2020), it was shown that if $v_{f,\bar{p}}(x)>0$, then
\[v_{f,\bar{p}}(x)=v_{f,\bar{p}}(f(\bar{p},v_{f,\bar{p}}(x)))\]
for all $x\in \Omega$, and $v_{f,\bar{p}}(0)=0$. Therefore, if $v_{f,\bar{p}}(x)>0$, then
\[u(x)=u(f(\bar{p},v_{f,\bar{p}}(x))),\]
and if $v_{f,\bar{p}}(x)=0$, then
\[u(x)=u(0).\]
If $v_{f,\bar{p}}(x)>0$, then for every sufficiently small $\varepsilon>0$, there exists $\delta>0$ such that if $y\in \Omega$ and $\|y-x\|<\delta$, then
\[u(f(\bar{p},v_{f,\bar{p}}(x)-\varepsilon))<u(y)<u(f(\bar{p},v_{f,\bar{p}}(x)+\varepsilon)),\]
which implies that $|v_{f,\bar{p}}(y)-v_{f,\bar{p}}(x)|<\varepsilon$. If $v_{f,\bar{p}}(x)=0$, then for every $\varepsilon>0$, there exists $\delta>0$ such that if $y\in \Omega$ and $\|y-x\|<\delta$, then
\[u(0)\le u(y)<u(f(\bar{p},\varepsilon)),\]
which implies that $0\le v_{f,\bar{p}}(y)<\varepsilon$. Therefore, $v_{f,\bar{p}}$ is continuous. This completes the proof. $\blacksquare$

\subsection{Proof of Theorem 1}
Define
\[I(t,c,g,p,q)=g((1-t)p+tq,c)\cdot (q-p),\]
where $g:\mathbb{R}^n_{++}\times \mathbb{R}_{++}\to \mathbb{R}^n$ is locally Lipschitz. Consider the following ODE:
\begin{equation}\label{eq6.4}
\dot{c}(t)=I(t,c(t),g,p,q),\ c(0)=m,
\end{equation}
and let $c(t;g,p,q,m)$ be the solution function. By Corollary 1, the domain of $c$ includes $[0,1]\times \{f^k\}\times \mathbb{R}^n_{++}\times \mathbb{R}^n_{++}\times \mathbb{R}_{++}$ for all $k$. First, we show that the domain of $c$ includes $[0,1]\times \{f\}\times \mathbb{R}^n_{++}\times \mathbb{R}^n_{++}\times \mathbb{R}_{++}$, and $\lim_{k\to \infty}c(t;f^k,p,q,m)=c(t;f,p,q,m)$ for all $t\in [0,1]$ and $(p,q,m)\in \mathbb{R}^n_{++}\times \mathbb{R}^n_{++}\times \mathbb{R}_{++}$.

Choose any $(p,m)\in \mathbb{R}^n_{++}\times \mathbb{R}_{++}$ and $q\in \mathbb{R}^n_{++}$, and define $p(t)=(1-t)p+tq$. For every continuous function $g:\mathbb{R}^n_{++}\times\mathbb{R}_{++}\to \mathbb{R}^n$, define
\[H_K(g)=\sup_{(r,c)\in [K^{-1},K]^{n+1}}\|g(r,c)\|.\]
Hereafter, we abbreviate $c(t;f,p,q,m)$ as $c(t)$ and $c(t;f^k,p,q,m)$ as $c^k(t)$. As we have already mentioned, $c^k(t)$ is defined on $[0,1]$ for all $k$ according to Corollary 1. If $p=q$, then our claim trivially holds. Hence, we assume that $p\neq q$.

Because $f^k$ is a demand function, there exists a weak order $\succsim_k$ on $\Omega$ such that $f^k=f^{\succsim_k}$. Define
\[E^k(r)=\inf\{r\cdot y|y\succsim_kf^k(p,m)\}.\]
By Fact 5, we have that $c^k(t)=E^k(p(t))$. Choose $w_0>0$ so small that $w_0<m$ and $nw_0p_i<q_im$ for all $i\in \{1,...,n\}$. If $q\cdot y\le w_0$, then $y_i\le w_0/q_i$, and thus
\[p\cdot y\le \sum_{i=1}^n\frac{w_0p_i}{q_i}<m.\]
Therefore, if $p(t)\cdot y=w_0$ for some $t\in [0,1]$, then $p\cdot y<m$, and thus $y\not\succsim^kf^k(p,m)$ for all $k$. By definition, $E^k(p(t))\ge w_0$, and thus $c^k(t)\ge w_0$ for all $k$ and $t\in [0,1]$. On the other hand,
\begin{align*}
c^k(t)=&~E^k(p(t))\le p(t)\cdot f^k(p,m)\\
\le&~(1-t)m+t\max\{q\cdot x|p\cdot x=m\}\\
\le&~m+\max\{q\cdot x|p\cdot x=m\}\equiv w_1
\end{align*}
for all $k$ and $t\in [0,1]$. Choose $K>1$ such that $p,q\in [K^{-1},K]^n$ and $w_0,w_1\in ]K^{-1},K[$. Because $f$ is locally Lipschitz, there exists $L>0$ such that if $(p',m'), (q',w')\in [K^{-1},K]^{n+1}$, then
\[\|f(p',m')-f(q',w')\|\le L\|(p',m')-(q',w')\|.\]
Let $I^*$ be the set of all $t\in [0,1]$ such that $c(s)$ is defined and $c(s)\in [K^{-1},K]$ for all $s\in [0,t]$. For any $t\in I^*$,
\begin{align*}
|c^k(t)-c(t)|\le&~\int_0^t\|f^k(p(s),c^k(s))-f(p(s),c(s))\|\|q-p\|ds\\
\le&~\int_0^t[\|f^k(p(s),c^k(s))-f(p(s),c^k(s))\|\\
&~+\|f(p(s),c^k(s))-f(p(s),c(s))\|]\|q-p\|ds\\
\le&~\int_0^tL\|q-p\||c^k(s)-c(s)|ds+H_K(f^k-f)\|q-p\|t,
\end{align*}
and thus, by Lemma 1,
\begin{equation}\label{eq6.5}
|c^k(t)-c(t)|\le \frac{H_K(f^k-f)}{L}(e^{L\|q-p\|}-1).
\end{equation}
This indicates that $c^k(t)\to c(t)$ as $k\to \infty$, and thus $c(t)\in [w_0,w_1]$ for all $t\in I^*$. Let $t^*=\sup I^*$. Because $c(t)$ is a nonextendable solution to (\ref{eq6.4}), Fact 3 implies that $c(t)$ is defined at $t^*$. By the continuity of $c(t)$, $c(t^*)\in [w_0,w_1]\subset [K^{-1},K]$, and thus $I^*=[0,t^*]$. If $t^*<1$, then $c(t^*)\in ]K^{-1},K[$, and thus there exists $t>t^*$ such that $t\in I^*$, which is a contradiction. Thus, $t^*=1$ and $I^*=[0,1]$, which implies that our claim holds.

Therefore, we have that the domain of the solution function $c$ includes $[0,1]\times \{f\}\times \mathbb{R}^n_{++}\times \mathbb{R}^n_{++}\times \mathbb{R}_{++}$. Fix $(p,m)\in \mathbb{R}^n_{++}\times \mathbb{R}_{++}$, and define $E(q)=c(1;f,p,q,m)$. By Lemma 3, $E$ solves (\ref{eq3.2}). By the above arguments, $E(q)=\lim_{k\to \infty}E^k(q)$ for all $q\in \mathbb{R}^n_{++}$. Because $E^k$ is concave, $E$ is also concave. By Corollary 1, $f$ is a demand function. This completes the proof. $\blacksquare$

\subsection{Proof of Corollary 3}
Suppose that $(f^k)$ is a sequence in $\mathscr{F}_L$. Let $p^*=(1,1,...,1)$ and $m^*=1$. Then, $(f^k(p^*,m^*))$ is a sequence on $[0,1]^n$. Therefore, it is bounded and there exists $M>0$ such that $\|f^k(p^*,m^*)\|\le M$ for all $k$. Moreover, this sequence has a convergent subsequence $(f^{\ell_1(k)}(p^*,m^*))$. Next, for $\nu\ge 2$, suppose that $\ell_{\nu-1}(k)$ is defined and $(f^{\ell_{\nu-1}(k)})$ is a uniformly convergent sequence on $\Delta_{\nu-1}$. Then, for any $(p,m)\in \Delta_{\nu}$,
\begin{align*}
\|f^{\ell_{\nu-1}(k)}(p,m)\|\le&~\|f^{\ell_{\nu-1}(k)}(p^*,m^*)\|+\|f^{\ell_{\nu-1}(k)}(p,m)-f^{\ell_{\nu-1}(k)}(p^*,m^*)\|\\
\le&~M+L_{\nu}\sqrt{n+1}\nu,
\end{align*}
which implies that $(f^{\ell_{\nu-1}(k)})$ is an equicontinuous and uniformly bounded sequence of functions on $\Delta_{\nu}$. By Ascoli--Arzel\`a's theorem, there exists a subsequence $(f^{\ell_{\nu}(k)})$ that uniformly converges on $\Delta_{\nu}$. Therefore, $(\ell_{\nu}(k))$ can be defined inductively. Define $\ell(k)=\ell_k(k)$. Then, $(f^{\ell(k)})$ is a subsequence of $(f^k)$ that converges to some function $f$ with respect to $\rho$. Clearly, $f$ is a continuous CoD that satisfies Walras' law. Moreover, if $(p,m), (q,w)\in \Delta_{\nu}$, then
\[\|f(p,m)-f(q,w)\|=\lim_{k\to \infty}\|f^{\ell(k)}(p,m)-f^{\ell(k)}(q,w)\|\le L_{\nu}\|(p,m)-(q,w)\|,\]
which implies that $f$ is locally Lipschitz, and by Theorem 1, $f\in \mathscr{F}_L$. This completes the proof. $\blacksquare$

\subsection{Proof of Theorem 2}
First, we show the following lemma.

\vspace{12pt}
\noindent
{\bf Lemma 4}. Suppose that $f$ is a locally Lipschitz demand function that satisfies Walras' law, and $R(f)$ includes $\mathbb{R}^n_{++}$. If $f$ satisfies the C axiom, then $u_{f,\bar{p}}$ is continuous on $\mathbb{R}^n_{++}$.\footnote{Note that, in this lemma, $R(f)$ need not be relatively open in $\Omega$, and thus Corollary 2 cannot be directly applied.}

\vspace{12pt}
\noindent
{\bf Proof}. Recall the differential equation (\ref{eq3.1}):
\[\dot{c}(t)=f((1-t)p+t\bar{p},c(t))\cdot (\bar{p}-p),\ c(0)=m.\]
Let $c(t;p,m)$ be the solution function. We have that $u_{f,\bar{p}}(x)=c(1;p,m)$ if $x=f(p,m)$. Choose any sequence $(x^k)$ in $\mathbb{R}^n_{++}$ such that $x^k\to x\in \mathbb{R}^n_{++}$ as $k\to \infty$, and suppose that $u_{f,\bar{p}}(x^k)\not\to u_{f,\bar{p}}(x)$. Taking a subsequence, we can assume that there exists $\varepsilon>0$ such that $|u_{f,\bar{p}}(x^k)-u_{f,\bar{p}}(x)|>\varepsilon$ for every $k$. Choose any $p^k\in G^f(x^k)$. Because $G^f$ is upper hemi-continuous, taking a subsequence, we can assume that $p^k\to p^*\in G^f(x)$. Then, 
\[u_{f,\bar{p}}(x^k)=c(1;p^k,p^k\cdot x^k)\to c(1;p^*,p^*\cdot x)=u_{f,\bar{p}}(x),\]
which is a contradiction. Therefore, $u_{f,\bar{p}}$ is continuous on $\mathbb{R}^n_{++}$. This completes the proof. $\blacksquare$

\vspace{12pt}
Choose any compact set $D\subset \mathbb{R}^n_{++}$. Let $x\in D$. We first show that there exist an open neighborhood $U$ of $x$ and $\varepsilon>0$ such that if $p\in G^{f^k}(y)$ for some $y\in U$ and $k$, then $p_i\ge \varepsilon$ for all $i\in \{1,...,n\}$.

Suppose not. Then, there exists a sequence $((p^{\ell},z^{\ell}))$ on $\mathbb{R}^n_{++}\times \Omega$ such that $p^{\ell}\in G^{f^{k(\ell)}}(z^{\ell})$ for all $\ell$, and $z^{\ell}\to x$ and $\min_jp^{\ell}_j\to 0$ as $\ell\to \infty$. First, suppose that $k(\ell)=k$ for infinitely many $\ell$. Taking a subsequence, we can assume that $k(\ell)=k$ for any $\ell$. By the C axiom, the inverse demand correspondence $G^{f^k}$ is compact-valued and upper hemi-continuous. Moreover, $p^{\ell}\in G^{f^k}(z^{\ell})$ for all $\ell$ and $z^{\ell}\to x$ as $\ell\to \infty$. Therefore, if we choose
\[\varepsilon=\frac{1}{2}\min\{\min_ip_i|p\in G^{f^k}(x)\},\]
then $\varepsilon>0$ and $\min_jp^{\ell}_j\ge \varepsilon$ for sufficiently large $\ell$, which is a contradiction. Hence, we can assume that $k(\ell)$ is increasing. Taking a subsequence, we can assume that $p^{\ell}\to p^*\in\mathbb{R}^n_+$, where $\sum_jp^*_j=1$ and $\min_ip^*_i=0$. Choose $i,j\in \{1,...,n\}$ such that $p^*_i>0$ and $p^*_j=0$.

Let $e_j$ denote the $j$-th unit vector. Choose a small $\delta>0$, and set $y_j=x_j+2$, $y_i(\delta)=x_i-\delta$, and $y_m=x_m$ for every $m\in \{1,...,n\}\setminus\{i,j\}$. Let $y(\delta)=(y_1,...,y_i(\delta),...,y_n)$. Because $G^f$ is upper hemi-continuous, there exists a sequence $(\delta_{\nu})$ of positive real numbers such that $\delta_{\nu}\to 0$ as $\nu\to \infty$ and there exists $p^{\nu}\in G^f(y(\delta_{\nu}))$ such that $p^{\nu}\to p^+\in G^f(y(0))$ as $\nu\to \infty$. Because $p^+\cdot y(0)>p^+\cdot (x+e_j)$, we have that $p^{\nu}\cdot y(\delta_{\nu})>p^{\nu}\cdot (x+e_j)$ for sufficiently large $\nu$. Choose such a $\nu$, and define $q=p^{\nu}$ and $y=y(\delta_{\nu})$. Then, $q\in G^f(y)$, and thus $y=f(q,q\cdot y)$. Define $y^k=f^k(q,q\cdot y)$. Then, $y^k\to y$ as $k\to \infty$. By assumption, $q\cdot z^{\ell}+q_j<q\cdot y$ for sufficiently large $\ell$, and thus, we have that $q\cdot z^{\ell}<q\cdot y^{k(\ell)}$ for sufficiently large $\ell$. However,
\[\lim_{\ell\to \infty}p^{\ell}\cdot z^{\ell}=p^*\cdot x>p^*\cdot y=\lim_{\ell\to \infty}p^{\ell}\cdot y^{k(\ell)},\]
and thus $p^{\ell}\cdot z^{\ell}>p^{\ell}\cdot y^{k(\ell)}$ if $\ell$ is sufficiently large, which contradicts the weak axiom of revealed preference for $f^{k(\ell)}$. Therefore, our initial claim is correct.

Define $U_x$ as such a neighborhood that corresponds with $x\in D$. Then, $(U_x)$ is an open covering of $D$, and thus, there exists $\varepsilon^*>0$ such that if $\sum_jp_j=1$ and $f^k(p,p\cdot x)=x$ for some $x\in D$ and $k$, then $p_i\ge \varepsilon^*$ for all $i\in \{1,...,n\}$.

Let
\[C=\{p\in\mathbb{R}^n_{++}|p\in G^f(x)\mbox{ for some }x\in D\},\]
\[C_k=\{p\in\mathbb{R}^n_{++}|p\in G^{f^k}(x)\mbox{ for some }x\in D\}.\]
By the compact-valuedness and upper hemi-continuity of inverse demand correspondences, $C,C_k$ are compact. Because of our previous arguments, there exists a compact set $K\subset \mathbb{R}^n_{++}$ that includes $C$ and all $C_k$. Define $m_1=\min\{p\cdot x|p\in K, x\in D\}>0$ and $m_2=\max\{p\cdot x|p\in K,x\in D\}>0$.

It suffices to show that $\sup_{x\in D}|u_{f^k,\bar{p}}(x)-u_{f,\bar{p}}(x)|\to 0$ as $k\to \infty$. Suppose not. Then, there exist $\varepsilon>0$ and a sequence $(x^{\ell})$ in $D$ such that $|u_{f^{k(\ell)},\bar{p}}(x^{\ell})-u_{f,\bar{p}}(x^{\ell})|\ge \varepsilon$ for all $\ell$, where $k(\ell)$ is increasing. Because $D$ is compact, we can assume that $x^{\ell}\to x^*\in D$ as $\ell\to \infty$. Suppose that $x^{\ell}=f^{k(\ell)}(p^{\ell},m^{\ell})$, where $p^{\ell}\in C_{k(\ell)}$ and $m^{\ell}=p^{\ell}\cdot x^{\ell}$. Taking a subsequence, we can assume that $p^{\ell}\to p^*\in K$. Define $m^*=p^*\cdot x^*$. Then, $(p^{\ell},m^{\ell}), (p^*,m^*)\in K\times [m_1,m_2]$, and thus,
\begin{align*}
\|f^{k(\ell)}(p^{\ell},m^{\ell})-f(p^*,m^*)\|\le&~\|f^{k(\ell)}(p^{\ell},m^{\ell})-f(p^{\ell},m^{\ell})\|\\
&~+\|f(p^{\ell},m^{\ell})-f(p^*,m^*)\|\to 0
\end{align*}
as $\ell\to \infty$. This implies that $f(p^*,m^*)=x^*$.

Now, consider the following differential equation:
\begin{equation}\label{eq6.6}
\dot{c}(t)=I(t,c(t),g,p,m),\ c(0)=m^*,
\end{equation}
where $g:\mathbb{R}^n_{++}\times \mathbb{R}_{++}\to \Omega$ is locally Lipschitz and $I(t,c,g,p,m)=g((1-t)p+t\bar{p},c+m-m^*)\cdot (\bar{p}-p)$. Let $c(t;g,p,m)$ be the solution function of (\ref{eq6.6}). We abbreviate $c(t;f,p^*,m^*)$ as $c^*(t)$ and $c(t;f^{k(\ell)},p^{\ell},m^{\ell})$ as $c_{\ell}(t)$. By Proposition 1, the domain of $c^*(t)$ and $c_{\ell}(t)$ includes $[0,1]$, $u_{f,\bar{p}}(x^*)=c^*(1)$, and $u_{f^{k(\ell)},\bar{p}}(x^{\ell})=c_{\ell}(1)+m^{\ell}-m^*$.

Choose $a>0$ and $b>0$ sufficiently small, and define
\[\Pi=\{(c,p,m)|\exists t\in [0,1]\mbox{ s.t. }|c^*(t)-c|\le a,\ \|p-p^*\|+|m-m^*|\le b\}.\]
Let $\hat{\Pi}$ be the set of all locally Lipschitz CoDs $g$ such that $\|g-f\|\le b$, where
\[\|h\|=\sup_{(t,c,p,m)\in [0,1]\times \Pi}\|h((1-t)p+t\bar{p},c+m-m^*)\|.\]
Define $\tilde{\Pi}=\Pi\times \hat{\Pi}$. We assume that $a,b$ are so small that 1) $\Pi$ is a compact set in $\mathbb{R}_{++}\times \mathbb{R}^n_{++}\times \mathbb{R}_{++}$, 2) $\min\{c+m-m^*|(c,p,m)\in \Pi\}>0$, 3) there exists $L>0$ such that if $(t.c,p,m), (t,c',p,m)\in [0,1]\times \Pi$, then 
\[\|f((1-t)p+t\bar{p},c+m-m^*)-f((1-t)p+t\bar{p},c'+m-m^*)\|\|\bar{p}-p\|\le L|c-c'|,\]
and 4) there exists $B>0$ such that if $(t,c,p,m,g)\in [0,1]\times \tilde{\Pi}$, then
\[|I(t,c,g,p,m)-I(t,c,f,p^*,m^*)|\le B[\|p-p^*\|+|m-m^*|+\|g-f\|].\]
Now, for any sufficiently large $\ell$, $\|p^{\ell}-p^*\|+|m^{\ell}-m^*|\le b$ and $\|f^{k(\ell)}-f\|\le b$. For such an $\ell$, define $t_{\ell}=\sup\{t\in [0,1]|\forall s\in [0,t], (c_{\ell}(s),p^{\ell},m^{\ell})\in \Pi\}$. Because $c_{\ell}(0)=m^*=c^*(0)$, we have that $t_{\ell}$ is well-defined and positive. If $t\in [0,t_{\ell}]$, then
\begin{align*}
&~|c_{\ell}(t)-c^*(t)|\\
\le &~\int_0^t|I(s,c_{\ell}(s),f^{k(\ell)},p^{\ell},m^{\ell})-I(s,c^*(s),f,p^*,m^*)|ds\\
\le &~\int_0^t|I(s,c_{\ell}(s),f^{k(\ell)},p^{\ell},m^{\ell})-I(s,c_{\ell}(s),f,p^*,m^*)|ds\\
&+\int_0^t|I(s,c_{\ell}(s),f,p^*,m^*)-I(s,c^*(s),f,p^*,m^*)|ds\\
\le&~\int_0^tL|c_{\ell}(s)-c^*(s)|ds+B(\|p^{\ell}-p^*\|+|m^{\ell}-m^*|+\|f^{k(\ell)}-f\|)s.
\end{align*}
By Lemma 1, we have that for any $t\in [0,t_{\ell}]$,
\begin{align*}
|c_{\ell}(t)-c^*(t)|\le&~\frac{B(\|p^{\ell}-p^*\|+|m^{\ell}-m^*|+\|f^{k(\ell)}-f\|)}{L}(e^L-1)\\
\equiv&~C(\|p^{\ell}-p^*\|+|m^{\ell}-m^*|+\|f^{k(\ell)}-f\|)
\end{align*}
for some $C>0$. Now, choose any $b'\in ]0,b]$ with $Cb'<a$. If $\ell$ is sufficiently large, then $\|p^{\ell}-p^*\|+|m^{\ell}-m^*|+\|f^{k(\ell)}-f\|\le b'$. For such an $\ell$, we have that $t_{\ell}=1$: if not, then $a\le |c_{\ell}(t_{\ell})-c^*(t_{\ell})|\le Cb'<a$, which is a contradiction. Therefore,
\[|c_{\ell}(1)-c^*(1)|\le C(\|p^{\ell}-p^*\|+|m^{\ell}-m^*|+\|f^{k(\ell)}-f\|),\]
and thus if $\ell$ is sufficiently large, then
\[|u_{f^{k(\ell)},\bar{p}}(x^{\ell})-u_{f,\bar{p}}(x^*)|\le |c_{\ell}(1)-c^*(1)|+|m^{\ell}-m^*|<\frac{\varepsilon}{2}.\]
By Lemma 4, $u_{f,\bar{p}}$ is continuous at $x^*$, and thus $|u_{f,\bar{p}}(x^{\ell})-u_{f,\bar{p}}(x^*)|<\frac{\varepsilon}{2}$ if $\ell$ is sufficiently large. Therefore,
\begin{align*}
\varepsilon \le&~|u_{f^{k(\ell)},\bar{p}}(x^{\ell})-u_{f,\bar{p}}(x^{\ell})|\\
\le&~|u_{f^{k(\ell)},\bar{p}}(x^{\ell})-u_{f,\bar{p}}(x^*)|+|u_{f,\bar{p}}(x^*)-u_{f,\bar{p}}(x^{\ell})|\\
<&~\varepsilon,
\end{align*}
which is a contradiction. This completes the proof. $\blacksquare$

\subsection{Proof of Corollary 4}
By Theorem 2, it suffices to show that $\limsup_{k\to \infty}v_{f^k,\bar{p}}(x)\le v_{f,\bar{p}}(x)$ for all $x\in \Omega\setminus \mathbb{R}^n_{++}$. Choose any $x\in \Omega\setminus \mathbb{R}^n_{++}$. Let $e=(1,1,...,1)$ and define $x^{\ell}=x+\ell^{-1}e$. By the same argument as in the proof of Corollary 2, we can show that $u_{f,\bar{p}}(x^{\ell})>v_{f,\bar{p}}(x)$ and $\lim_{\ell\to \infty}u_{f,\bar{p}}(x^{\ell})=v_{f,\bar{p}}(x)$. The same fact is true for $v_{f^k,\bar{p}}$. Choose $\varepsilon>0$. Then, there exists $\ell$ such that $u_{f,\bar{p}}(x^{\ell})<v_{f,\bar{p}}(x)+\varepsilon$. Because $u_{f^k,\bar{p}}(x^{\ell})\to u_{f,\bar{p}}(x^{\ell})$, we have that $u_{f^k,\bar{p}}(x^{\ell})<u_{f,\bar{p}}(x^{\ell})+\varepsilon$ for sufficiently large $k$. Therefore, for such $k$, $v_{f^k,\bar{p}}(x)<u_{f^k,\bar{p}}(x^{\ell})<v_{f,\bar{p}}(x)+2\varepsilon$. Hence, $\limsup_{k\to \infty}v_{f^k,\bar{p}}(x)\le v_{f,\bar{p}}(x)+2\varepsilon$, and because $\varepsilon>0$ is arbitrary, $\limsup_{k\to \infty}v_{f^k,\bar{p}}(x^{\ell})\le v_{f,\bar{p}}(x)$, as desired. This completes the proof. $\blacksquare$

\subsection{Proof of Corollary 5}
By Theorem 2, it suffices to show that if $(f^k)$ is a sequence on $\mathscr{F}_{L,M}$ that converges to $f\in \mathscr{F}_L$ with respect to $\rho$, then $f\in \mathscr{F}_{L,M}$.

First, choose any $x\in \mathbb{R}^n_{++}$. Then, there exists $\nu$ such that $x\in ]\nu^{-1},\nu[^n$. Choose $p^k\in G^{f^k}(x)$ for each $k$. Because $(p^k)$ is a sequence of the compact set
\[P_{\nu}=\{p\in \mathbb{R}^n_{++}|\sum_ip_i=1,\ \min_ip_i\ge M_{\nu}\},\]
there exists a subsequence $p^{\ell(k)}$ such that $\lim_{k\to \infty}p^{\ell(k)}=p^*\in P_{\nu}$. Because $f^{\ell(k)}$ converges to $f$ uniformly on any compact set,
\[x=\lim_{k\to \infty}f^{\ell(k)}(p^{\ell(k)},p^{\ell(k)}\cdot x)=f(p^*,p^*\cdot x),\]
and thus $p^*\in G^f(x)$. This implies that $R(f)$ includes $\mathbb{R}^n_{++}$.

Second, choose any $x\in ]\nu^{-1},\nu[^n$, and suppose that there exists $p\in G^f(x)$ such that $\min_ip_i<M_{\nu}$. Let $x^k=f^k(p,p\cdot x)$. Then, $x^k\to x$ as $k\to \infty$, and thus $x^k\in ]\nu^{-1},\nu[^n$ for sufficiently large $k$. Because $p\in G^{f^k}(x^k)$, we have that $f^k\notin \mathscr{F}_{L,M}$, which is a contradiction. Therefore, if $p\in G^f(x)$, then $\min_ip_i\ge M_{\nu}$. Because $G^f(x)$ is obviously closed, this implies that $G^f$ is compact-valued.

Third, it is easy to show that for any demand function $f'$ and $x\in R(f')$, $G^{f'}(x)$ is convex. Therefore, $G^f$ is convex-valued.

Finally, suppose that $G^f$ is not upper semi-continuous at $x$. Then, there exist an open neighborhood $U$ of $G^f(x)$ and sequences $(x^{\ell})$ and $(p^{\ell})$ such that $x^{\ell}\to x$ as $\ell\to \infty$ and $p^{\ell}\in G^f(x^{\ell})\setminus U$ for all $\ell$. Choose $\nu$ such that $x^{\ell}\in ]\nu^{-1},\nu[^n$ for all $\ell$. Then, $(p^{\ell})$ is a sequence in the compact set $P_{\nu}$. Thus, by taking a subsequence, we can assume that $p^{\ell}\to p^*\in P_{\nu}$ as $\ell\to \infty$. Because $f$ is continuous, we have that $p^*\in G^f(x)\subset U$, which is a contradiction. Therefore, $f$ satisfies the C axiom. Hence, $f\in \mathscr{F}_{L,M}$, as desired. This completes the proof. $\blacksquare$

\subsection{Proof of Theorem 3}
By Corollary 3, $\mathscr{F}_L$ is compact with respect to $\rho$. Therefore, there exists a subsequence $(f^{\ell(k)})$ of $(f^k)$ such that for some $g\in \mathscr{F}_L$, $\rho(f^{\ell(k)},g)\to 0$ as $k\to \infty$. Because $(f^k)$ converges to $f$ pointwise, we have that $f=g$, and thus $f\in \mathscr{F}_L$.

Next, suppose that $f^k\in \mathscr{F}_{L,M}$ for any $k$. By Corollary 5 and the same argument as in the above paragraph, we have that $f\in \mathscr{F}_{L,M}$. Suppose that for some compact set $D\subset \mathbb{R}^n_{++}$,
\[\limsup_{k\to \infty}\sup_{x\in D}|u_{f^k,\bar{p}}(x)-u_{f,\bar{p}}(x)|>0.\]
Taking a subsequence, we can assume that there exists $\varepsilon>0$ such that $\sup_{x\in D}|u_{f^k,\bar{p}}(x)-u_{f,\bar{p}}(x)|\ge \varepsilon$ for all $k$. Because $\mathscr{F}_L$ is compact with respect to $\rho$, there exists a subsequence $(f^{\ell(k)})$ such that $\lim_{k\to \infty}\rho(f^{\ell(k)},f)=0$. By Corollary 5, for any sufficiently large $k$, $\sup_{x\in D}|u_{f^{\ell(k)},\bar{p}}(x)-u_{f,\bar{p}}(x)|<\varepsilon$, which is a contradiction. This completes the proof. $\blacksquare$

\section*{References}
\begin{description}
\item{[1]} Allen, R., Dziewulski, P., Rehbeck, J.: Revealed statistical consumer theory. Econ. Theory (2023)

\item{[2]} Antonelli, G. B.: Sulla Teoria Matematica dell' Economia Politica. Tipografia del Folchetto, Pisa (1886). Translated by Chipman, J. S., Kirman, A. P. 1971. On the mathematical theory of political economy. In: Chipman, J. S., Hurwicz, L., Richter, M. K., Sonnenschein, H. F. (Eds.) Preferences, Utility and Demand, pp.333-364. Harcourt Brace Jovanovich, New York (1971)

\item{[3]} Blundell, R., Horowitz, J. Parey, M.: Nonparametric estimation of a nonseparable demand function under the Slutsky inequality restriction. Rev. Econ. Stat. 99, 291-304 (2017)

\item{[4]} Deaton, A.: Demand analysis. In: Griliches, Z., Intriligator, M. D., (Eds.) Handbook of Econometrics, Vol.3, pp.1767-1839. Elsevier, Amsterdam (1986)

\item{[5]} Debreu, G.: Representation of a preference ordering by a numerical function. In: Thrall, R. M., Coombs, C. H., Davis, R. L. (Eds.) Decision Processes, pp.159-165. Wiley, New York (1954)

\item{[6]} Debreu, G.: Smooth preferences. Econometrica 40, 603-615 (1972)

\item{[7]} Debreu, G.: Excess demand functions. J. Math. Econ. 1, 15-21 (1974)

\item{[8]} Dieudonne, J.: Foundations of Modern Analysis. Academic Press, London (1969)

\item{[9]} Hartman, P.: Ordinary Differential Equations. second ed. Birkh\"auser Verlag AG, Boston (1997)

\item{[10]} Heinonen, J.: Lectures on Lipschitz analysis. Lectures at the 14th Jyv\"askyl\"a Summer School (2004)

\item{[11]} Hosoya, Y.: Measuring utility from demand. J. Math. Econ. 49, 82-96 (2013)

\item{[12]} Hosoya, Y.: A Theory for estimating consumer's preference from demand. Adv. Math. Econ. 18, 33-55 (2015)

\item{[13]} Hosoya, Y.: The relationship between revealed preference and the Slutsky matrix. J. Math. Econ. 70, 127-146 (2017)

\item{[14]} Hosoya, Y.: First-order partial differential equations and consumer theory. Discret. Contin. Dyn. Syst. - Ser. S 11, 1143-1167 (2018)

\item{[15]} Hosoya, Y.: Recoverability revisited. J. Math. Econ. 90, 31-41 (2020)

\item{[16]} Hosoya, Y.: Consumer optimization and a first-order PDE with a non-smooth system. Oper. Res. Forum 2:66 (2021)

\item{[17]} Houthakker, H. S.: Revealed preference and the utility function. Economica 17, 159-174 (1950)

\item{[18]} Hurwicz, L.: On the problem of integrability of demand functions. In: Chipman, J. S., Hurwicz, L., Richter, M. K., Sonnenschein, H. F. (Eds.) Preferences, Utility and Demand, pp.174-214. Harcourt Brace Jovanovich, New York (1971)

\item{[19]} Hurwicz, L., Uzawa, H.: On the integrability of demand functions. In: Chipman, J. S., Hurwicz, L., Richter, M. K., Sonnenschein, H. F. (Eds.) Preferences, Utility and Demand, pp.114-148. Harcourt Brace Jovanovich, New York (1971)

\item{[20]} Ioffe, A. D., Tikhomirov, V. M.: Theory of Extremal Problem. North Holland, Amsterdam (1979)

\item{[21]} Karatzas, I, Shreve, S. E.: Brownian Motion and Stochastic Calculus: 2nd ed. Springer, New York (1998)

\item{[22]} Katzner, D. W.: Static Demand Theory. Macmillan, London (1970)

\item{[23]} Mas-Colell, A.: The recoverability of consumers' preferences from market demand behavior. Econometrica 45, 1409-1430 (1977)

\item{[24]} Mas-Colell, A.: The Theory of General Economic Equilibrium: a Differentiable Approach. Cambridge University Press, Cambridge (1985)

\item{[25]} Mas-Colell, A., Whinston, M. D., Green, J.: Microeconomic Theory. Oxford University Press, Oxford (1995)

\item{[26]} Nikliborc, W.: Sur les \'equations lin\'eaires aux diff\'erentielles totales. Studia Mathematica 1, 41-49 (1929)

\item{[27]} Pareto, V.: Manuale di Economia Politica con una Introduzione alla Scienza Sociale. Societa Editrice Libraria, Milano (1906)

\item{[28]} Richter, M. K.: Revealed preference theory. Econometrica 34, 635-645 (1966)

\item{[29]} Rudin, W.: Real and Complex Analysis, third ed. McGraw-Hill, Singapore (1987)

\item{[30]} Samuelson, P. A.: The problem of integrability in utility theory. Economica 17, 355-385 (1950)

\item{[31]} Shiozawa, K.: Revealed preference test and shortest path problem; graph theoretic structure of the rationalization test. J. Math. Econ. 67, 38-48 (2016)

\item{[32]} Uzawa, H.: Preferences and rational choice in the theory of consumption. In: Arrow, K. J., Karlin, S., Suppes, P. (Eds.) Mathematical Methods in the Social Sciences, 1959: Proceedings of the First Stanford Symposium, pp.129-149. Stanford University Press, Stanford (1960) Reprinted in: Chipman, J. S., Hurwicz, L., Richter, M. K., Sonnenschein, H. F. (Eds.) Preferences, Utility, and Demand, pp.7-28. Harcourt Brace Jovanovich, New York (1971)

\end{description}

\end{document}